\documentclass[structabstract]{aa}  
\usepackage{epsfig}          
\usepackage{color}     
\usepackage{natbib}        
\usepackage{txfonts}
\graphicspath{{Plot_final/}}
\newcommand{\tst}{t_{\rm onset}}

\usepackage{graphicx}
\begin{document}

   \title{Interplanetary magnetic structure guiding solar relativistic particles}

   \author{S.~Masson\inst{1,2}, P.~D\'emoulin\inst{1}, S.~Dasso\inst{3}, and K.-L.~Klein\inst{1}}
   \institute{  \inst{1} LESIA, Observatoire de Paris, CNRS, UPMC, Universit\'e Paris Diderot, 5 place Jules Janssen, 92190 Meudon, France. \\
   \inst{2} presently at Space Weather Laboratory, NASA- Goddard Space Flight Center, 8800 Greenbelt Road, Greenbelt, MD 20771, USA \\
\email{sophie.masson@nasa.gov}\\
     \inst{3} Instituto de Astronom\'ia y F\'isica del Espacio (CONICET-UBA) and 
Departamento de F\'isica, Facultad de Ciencias Exactas y Naturales (UBA), Buenos Aires, Argentina
}

   \date{Received September 26, 2011}

\abstract
{Relating in situ measurements of relativistic solar particles to their parent activity in the corona requires understanding the magnetic structures that guide them from their acceleration site to the Earth. Relativistic particle events are observed at times of high solar activity, when transient magnetic structures such as Interplanetary Coronal Mass Ejections (ICMEs) often shape the interplanetary magnetic field (IMF). They may introduce interplanetary paths that are longer than nominal, and magnetic connections rooted far from the nominal Parker spiral.
}
{We present a detailed study of the IMF configurations during ten relativistic solar particle events of the 23rd activity cycle to elucidate the actual IMF configuration guiding the particles to Earth, where they are measured by neutron monitors.
}
{We use magnetic field  (MAG) and plasma parameter measurements (SWEPAM) from ACE, and determine interplanetary path lengths of energetic particles through a modified version of the velocity dispersion analysis based on energetic particle measurements with SoHO/ERNE. 
}
{We find that the majority (7/10) of the events is detected in the vicinity of an ICME. Their interplanetary path lengths are found to be longer (1.5-2.6~AU) than those of the two events propagating in the slow solar wind (1.3~AU). The largest apparent path length is found in an event within the fast solar wind, probably due to enhanced pitch angle scattering. The derived path lengths imply that the first energetic and relativistic protons are released at the Sun at the same time as
electron beams emitting  type III radio bursts.
}
{The timing of the first high-energy particle arrival at Earth is dominantly determined by the type of IMF in which the particles propagate. Initial arrival times are as expected  from Parker's model in the slow solar wind, and significantly larger in or near transient structures such as ICMEs.
} 

\titlerunning{Interplanetary magnetic structure guiding solar relativistic particles}
  \authorrunning{Masson et al.}

  \maketitle

\section{Introduction}  
 \label{Introduction}

The solar activity has consequences for the entire inner heliosphere.  Besides high-energy photons, one can distinguish two energetic phenomena directly affecting the terrestrial environment: coronal mass ejections (CMEs), disturbing  the magnetized environment, and the energetic particle events, impacting the Earth's atmosphere and affecting the ionized environment.  The most energetic particles that the Sun produces are relativistic protons up to about 10~GeV. At these relativistic energies, solar particles penetrate the magnetosphere and impact the Earth's atmosphere. The atmospheric interaction of the GeV particles produces secondary particles through nuclear cascades that are detected at the ground level by neutron monitors (NMs). Hence, the name of ground level enhancement or GLE. Only 70 events have been reported since 1942, and have been largely studied in order to address the acceleration and propagation of these particles from Sun to the Earth. 

Solar energetic phenomena may accelerate particles up to relativistic energies through the coronal shock driven by the CME \citep{VianioLethi07,SandroosVainio09}, or through magnetic reconnection during a flare \citep{ArznerVlahos04,Dmitruk_al04,Turkmani_al05,Drake_al06}.
 
 Energetic particles are guided by the interplanetary magnetic flux tube connecting the acceleration site to the Earth. Most previous studies are based on the assumption that energetic particles propagate along the nominal Parker spiral.  Nevertheless, two main problems arise under this assumption.  First, it implies that the parent active regions are located near $30\degr-80\degr$ West. However,  observations show  that the distribution in longitude of the parent active regions is  broad, ranging from $90\degr$ East to more than $120\degr$ West  \citep{Cliver_al82,Kahler_al84}. Second, the time measured between the radiative signatures of particle acceleration in the Sun's corona and the in-situ measurements is longer than it should be for a propagation in the Parker spiral  \citep{Debrunner_al97,Kahler_al03,Tylka_al03}.

Several scenarios have been proposed to explain both the delay and the connection problem during relativistic particle events. Indeed, the delay may be due to a late acceleration and/or injection phase during the flare \citep{Debrunner_al97,Klein_al99} or to a time extended acceleration at the bow shock of the CME \citep{Reames99,Gopal05}.  Since shock acceleration should inject particles over a large angular width, explaining also the connection problem, several studies favored the CME shock acceleration rather than the flare acceleration \citep{Cliver82,HudLinStew82,Cliver_al04}.

However, it has been known since early investigations of relativistic solar energetic particle events that they often occur during depressions of the galactic cosmic ray intensity. Such a depression is typically associated with a transient interplanetary magnetic field structure \citep[see][ and references therein]{Carmichael62}, now known as an interplanetary coronal mass ejection (ICME), which is the interplanetary counter part of the CME observed in the solar corona \citep{Wimmer_al06}.  While observed at 1~AU, ICMEs can still be connected at the Sun by one or two footpoints \citep[e.g.][]{Crooker_al08}.  Thus, relativistic particles could be detected at Earth in ICMEs when the parent active regions is far from the nominally well-connected western solar hemisphere \citep{Debrunner_al88}. Detailed studies of individual events showed indeed that energetic particles can propagate within ICMEs \citep{RichCaneRosen91,Larson_al97,TorRiiKoch04}.

Moreover, the specific magnetic topology of ICMEs modifies the interplanetary propagation of energetic particles such as the directivity of the particles flux \citep{Ruffolo_al09}. This could explain peculiarities of directional distributions of relativistic protons observed at Earth \citep{Bieber_al02,Miroshnichenko_al05,Ruffolo_al06,Saiz_al08}.  Also, long geometrical paths in the interplanetary space inferred from velocity dispersion analysis \citep{Larson_al97} or from detailed timing comparisons \citep{Masson_al09a} point to particle propagation in non-nominal interplanetary magnetic fields. 
 
Thus, the relationship between solar energetic particles measured near 1~AU and the parent activity in the corona depends on a combination of processes: particle acceleration in the corona, access of the accelerated particles to open field lines, and propagation through interplanetary space along various possible magnetic field configurations. 

In this paper we explore the interplanetary magnetic structures and their impact on particle propagation during the most energetic solar particle events, the GLEs. Our analysis is based on two independent methods: the identification of the interplanetary magnetic structures using magnetic field and plasma parameters 
measured aboard the ACE spacecraft (Section~\ref{s-imf}) and the velocity dispersion analysis of the initial proton arrival times based on ERNE/SoHO and neutron monitor data (Section~\ref{s-vda}). 
This is completed by comparing the deduced solar release time of protons with the time interval of electrons as deduced from type III  bursts observations. Then we summarize and conclude in Section~\ref{Conclusion}.

\section{The interplanetary magnetic structures guiding relativistic particles} 
\label{s-imf}

\subsection{Characteristics of interplanetary structures} 
\label{s-imf-Characteristic}

The magnetic field vector at the spacecraft is conveniently characterized by the magnitude of the field, $B$, its orientation defined by the latitude, $\theta_B$, above the ecliptic plane and the longitude, $\phi_B$, with respect to the Sun-Earth axis. In the case of the undisturbed solar wind, where the field line is a Parker Spiral, the magnetic field is weak ($B\sim 5~\rm{nT}$ at 1~AU) and strongly non coherent, with angles fluctuating around typical mean values $\theta_B \simeq 0\degr$ and $\phi_B \simeq -45\degr$ or $135~\degr$ (for an inward or outward sector, respectively, with GSE coordinates). When an ICME, initially ejected from the solar corona, reaches the spacecraft, $B$ increases strongly and it becomes more coherent \citep[e.g.,][]{Burlaga95}. The magnetic field orientation evolves also more coherently. 

In the solar wind, the temperature of protons is empirically dependent on the proton velocity \citep{LopezFreeman81}. \citet{Elliott_al05} established the relation 
   \begin{equation}
   T_{\rm exp}\rm{[K]}=640\times V_{\rm SW}[\rm{km.s}^{-1}]-1.56\times 10^{5}
   \end{equation}

\noindent between the expected proton temperature, $T_{\rm exp}$, and the solar wind speed, $\rm{V}_{\rm SW}$. A deviation from this relationship allows us to identify a transient interplanetary structure. An ICME has typically a proton temperature lower than $T_{\rm exp}/2$ (\citealt{Burlaga_al81}, \citealt{KleinBurl82}, \citealt{Burlaga91}), and this criterion is frequently used to define the extent of an ICME \citep[][and references therein]{RichCane10}. Moreover, if the magnetic field is also significantly more intense than in the solar wind, with a coherent rotation, this region defines a magnetic cloud. 

Differences between solar wind and ICMEs are also expected in the proton $\beta$ (ratio of proton over magnetic pressure, noted $\beta_p$). Typically $\beta_p \ge 0.4$ in the solar wind, while it is generally smaller, typically in the range $ 0.01 \le \beta_p \le 0.4 $, within an ICME or a magnetic cloud, which are cooler and have stronger magnetic field.

When a magnetic cloud (or an ICME) moves faster than the surrounding solar wind, a sheath is formed in front with enhanced plasma density and magnetic field strength. Magnetic reconnection is typically expected between two magnetic structures when different magnetic fields are pushed against each other. In situ evidences of such reconnection have recently been found in a magnetic cloud sheath \citep{Chian11}. More generally, reconnection between the sheath and flux rope magnetic field implies that the flux rope front can be progressively pealed, while at the flux rope rear the corresponding magnetic flux region remains (called a back region).  Because of the different magnetic topologies, the back region is typically separated from the remnant flux rope by a current sheet.  This back region has intermediate properties between those of the solar wind and the magnetic cloud, since it originally belongs to the flux rope, but after reconnection it is also connected to the solar wind \citep{Dasso_al06,Dasso_al07}.  

A priori, solar energetic particles can travel in any of the interplanetary structures summarized above.  So, depending on their injection site on the Sun, or their access to the structure, energetic particles could be observed in a parkerian solar wind or in an environment associated with an ICME (e.g., in the ICME sheath, inside a magnetic cloud flux rope, in the magnetic cloud back, or in the distorted solar wind that is perturbed by the fast transient).

\begin{table*}
\caption[] {Summury of the studied relativistic events}\label{t-event}
\center{
\begin{tabular}{c @{ } c c c c c c @{} c  }
\hline
Column 1                          &  2             &   3          &     4       &     5                 &  6       &7       & 8 \\  
\# GLE~/ date                   &$\tst$(UT)& Max. \% & Active region& PS                 & $V_{\rm SW}$           &    Connection    &Interplanetary  \\
                                                    &       &  increase &  location      &     footpoints     &	(km.s$^{-1}$)	      &                          &  structure      \\ 
\hline
59 / 14 {\tt Jul.} 00~~               &10:31&     59       &N22,W07     &W43 & 500 &poorly& back region  \\ 
60 / 15 {\tt Apr.} 01$^{\star}$  &13:55&   237       &S20,W85     &W33 & 640 &poorly& disturbed solar wind \\ 
61 / 18 {\tt Apr.} 01$^{\star}$  &02:36&     26       & behind limb &W60& 360 &poorly& ICME sheath  \\ 
63 / 26 {\tt Dec.} 01$^{\star}$ &05:40&     13       & N08, W54   &W51 & 420 & well & slow solar wind  \\
64 / 24 {\tt Aug.} 02$^{\star}$ &01:24&     14       & S02, W81   &W54 & 400 & well & slow solar wind   \\ 
65 / 28 {\tt Oct.} 03$^{\star}$  &11:12&     47       &S16, E08   &W48 & 450   &poorly&  back region     \\
66 / 29 {\tt Oct.} 03~~	               &21:01&     35       & S15, W02   &W48   & 450             &poorly& magnetic cloud \\
67 / 02 {\tt Nov.} 03$^{\star}$ &17:30&     39       &S14, W56   &W36 & 600 & well & back region      \\
69 / 20 {\tt Jan.} 05~~	     &06:49& 5400       &N14, W55   &W39 & 550 & well & disturbed solar wind \\ 
70 / 13 {\tt Dec.} 06$^{\star}$ &02:50&      92       & S06, W23   &W33 & 660 & well & fast solar wind \\ 
\hline
\end{tabular}
}
\begin{list}{}{}
 \item[{\it Note :}] Results obtained from independent studies are synthesized in this table. 
{\it Col.~1}: the number and the date of the GLEs. 
{\it Col.~2}: the onset time of the first responding Neutron Monitor. 
{\it Col.~3}: Maximum percentage increase of the GLE \citep{Belov_al10}.
{\it Col.~4}: the location of the parent active region.
{\it Col.~5}: the solar longitude of the Parker spiral footpoint. 
{\it Col.~6}: The mean velocity of the solar wind computed on a half-day time interval for a quiet region of the IMF immediately  preceding any transient magnetic structures transiting through the Earth during the GLE.  
{\it Col.~7}: The nature of connection between the parent active region and the Parker spiral footpoint.
{\it Col.~8}: results of the identification of the magnetic topology of the interplanetary medium. 
The stars on the right of {\it Col. 1} indicate the GLEs where interplanetary length and solar release time could be estimated (Table~\ref{t-length}).
\end{list}
\end{table*}

\subsection{GLE~selection and characterization of the interplanetary magnetic structures}
\label{s-imf-Principle}

We initially select the relativistic events reported between 2000 and 2006. We do not study the GLEs with a marginal increase of the neutron monitor count rate ($<5\%$). The list is thus restrained to 10 events (Table~\ref{t-event}). The arrival times of the first detected protons at Earth have been already published by \citet{Moraal_al11}, and we report their results in Table~\ref{t-event}. 

Our analysis is based on measurements of the plasma and magnetic field near the L1 Lagrange point of the Sun-Earth system by the {\it Advanced Composition Explorer} (ACE) mission \citep{Stone_al98}:  the magnetic data are obtained from the MAG instrument \citep{Smith_al98} and the plasma data from the SWEPAM instrument \citep{McComas_al98}. In order to compare with neutron monitor measurements at 1~AU, we have to propagate the ACE measurements to the Earth. We perform this time correction by assuming bodily propagation of the magnetic field structures from L1 to Earth at the average solar wind proton speed $V_p$ measured with ACE / SWEPAM during the two hours preceding the arrival of protons at Earth.  This implies a time shift $ \Delta t = D_{L1}/V_p$ where $D_{L1}$ is the L1-Earth distance.

\begin{figure}
 \hspace{-0.7cm}  
   \centering
   \includegraphics[width=0.515\textwidth, clip=]{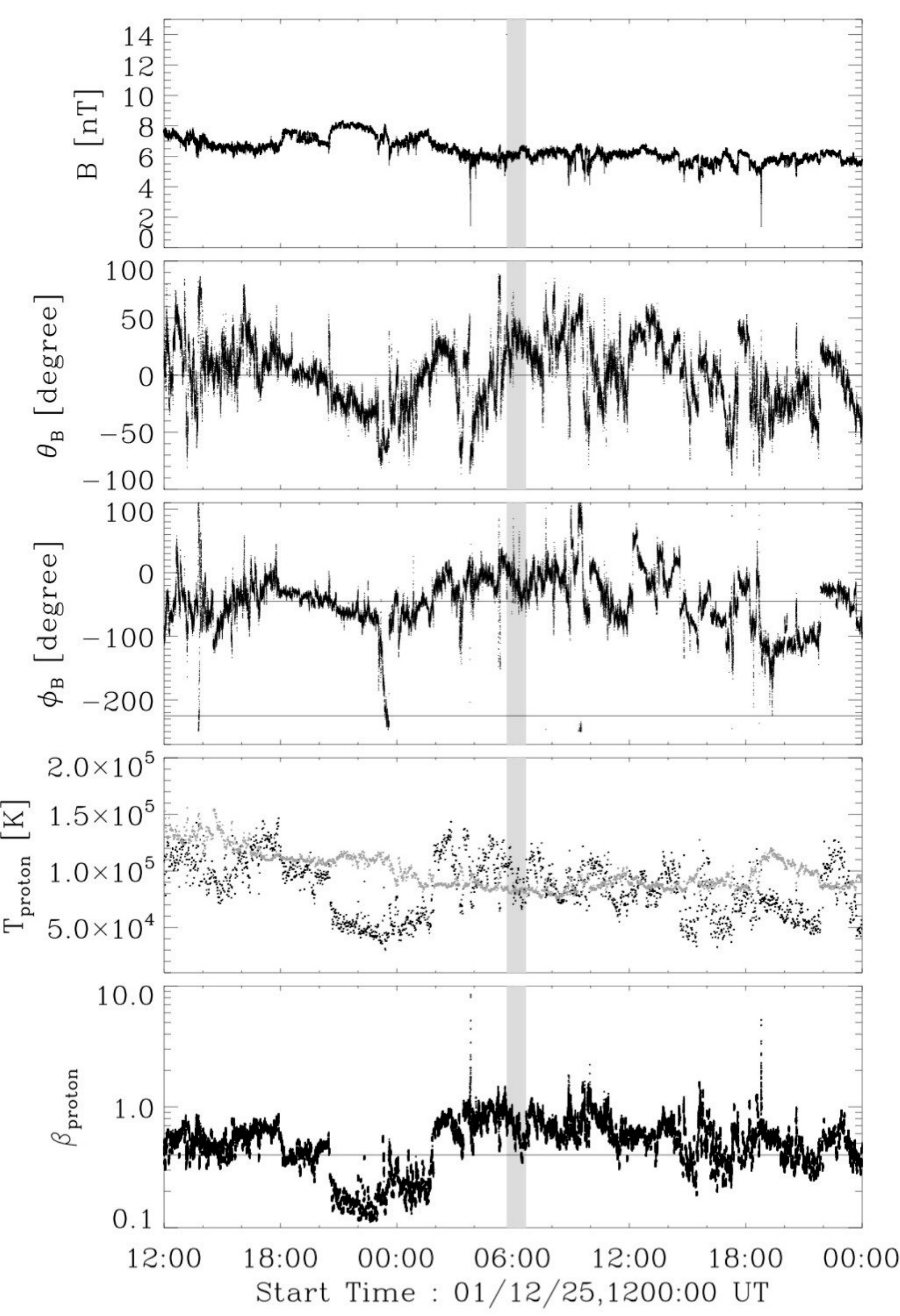}
   \caption{Interplanetary data around GLE~63 (Dec. 26, 2001): example of a GLE~present in a typical Parker spiral. {\it From the top to the bottom}, panels display respectively: the magnitude of the magnetic field ($\rm{nT}$); its latitude  $\theta_B$ and its longitude $\phi_B$ (given in GSE coordinate system); then on the same panel are over plotted the temperature of protons measured by the spacecraft ({\it black dots}) and the expected temperature ({\it gray dots}), and finally the $\beta$ of protons. An horizontal line is added for $\theta_B=0$, $\phi_B=-45\degr$ and $\beta_{\rm proton}=0.4$ to mark characteristic values.
The grey-scale rectangle on the 5 panels corresponds to the 1-hour time interval starting at the first arrival time of relativistic protons at Earth. The time axis is as measured at Earth. ACE measurements at the L1 Lagrangian point have been shifted by the appropriate travel time (see Section \ref{s-imf-Principle}).
           }
  \label{fig1}
 \end{figure}

The magnetic field and the plasma parameters are drawn on adjacent panels, in order to compare their temporal evolution (e.g. Figs.~\ref{fig1}-\ref{fig3}).  By over-plotting as a grey bar the 1-hour-time interval starting with the first arrival time of relativistic particles at the first responding neutron monitor, we characterize the magnetic structure in which the first relativistic protons arrive at the Earth.

We localize their parent active region ({\it col.~4} of Table~\ref{t-event}) and the longitude of the footpoint of the Earth-connected Parker spiral field line, given in {\it col.~5} of Table~\ref{t-event}. The longitude of the Parker spiral root on the solar source surface is at \citep{Parker61}
   \begin{equation}
    \Phi_S = \Phi(1\rm{~AU})+\frac{\Omega}{V_{\rm SW}}(1~\rm{AU}-R_S)  \,,
    \label{eq-parker}
   \end{equation}
where $\Omega=2.6\times 10^{-6}~\rm{rad.s}^{-1}$ is the angular speed of the Sun, 
$V_{\rm SW}$ the solar wind speed, assumed constant from the Sun to the Earth, $R_s=2.5 R_{\odot}$ is the radius of the spherical source surface, and $\Phi(1\rm{~AU})$ is the longitude of the spiral at 1~AU.  We select $\Phi(1\rm{~AU})=0$ as the reference longitude and $\Phi_S$ is positive westward from the Sun-Earth axis. 
Because of magnetic fluctuations, \citet{Ippolito_al05} showed by a numerical study that the longitude of the Parker spiral footpoints at the source surface can not be estimated with an accuracy better than $6\degr - 10\degr$. Moreover, coronal magnetic field extrapolations showed that an open coronal flux tube rooted in an active region can spread by several tens of degrees in longitude westward and eastward \citep{Klein_al08}. Therefore, we consider that the parent active region is well-connected to the Earth by the Parker spiral if its longitude is within the range $[\Phi_S-30\degr;\Phi_S+30\degr]$, otherwise we define the parent active region as being poorly connected ({\it col.~7} of Table~\ref{t-event}). The path lengths along the nominal Parker spirals in this sample range from 1.06~AU ($V_{SW}=660 \rm km s^{-1}$) to 1.15~AU ($400~$\rm km.s$^{-1}$),  {\it i.e.} they are essentially 1.1~AU for each event.

\subsection{Magnetic configurations guiding relativistic particles}
\label{s-imf-interplanetary}

Our analysis leads us to separate the magnetic field configurations along which relativistic particles reached the Earth into three distinct subsets ({\it col. 8} of Table~\ref{t-event}): the nominal solar wind with a Parker spiral field (Section~\ref{s-imfps}), a transient magnetic structure related to an ICME (Section~\ref{s-imfmc}), and a third category, typically a highly disturbed solar wind in the vicinity of an ICME (Section~\ref{s-imfcompl}).

\subsubsection{The Parker spiral} \label{s-imfps}
 
In our sample, only 3 GLEs reach the Earth along a typical Parker spiral: Dec. 26, 2001 (GLE~63), Aug. 24, 2002 (GLE~64) and Dec. 13, 2006 (GLE~70). 

Between 02:00 - 24:00~UT on Dec. 26, 2001, the magnitude of the magnetic field is fluctuating around $5-6$~nT (Figure~\ref{fig1}). The magnetic field orientation is typical of the Parker spiral ($\theta_{B} \approx 0\degr $ and $\phi_{B} \approx -45\degr$). Moreover, the expected temperature is almost the same as  the observed one and $0.4 \leq \beta_P \leq 1$. 
A small magnetic cloud could be present between 21:00 UT on Dec. 25 and 02:00 UT on Dec. 26, but it was not listed by \citet{RichCane10} and it ends about 4 hours before the arrival of relativistic particles.
Thus, the relativistic particles impacting the Earth at 05:40~UT on Dec. 26, 2001 (GLE~63), propagate along the Earth-connected Parker field line.

The two other events consistent with a propagation along the nominal Parker spiral are GLE~64 and GLE~70. The magnetic field orientation, the plasma beta and the similarity of the expected and the actually observed temperature all clearly argue in favor of this interpretation 
(see Sections~\ref{A-GLE64}, \ref{A-GLE70} and Figures~\ref{figA1-1}, \ref{figA1-2}).

For these 3 events, no ICMEs have been identified few days before the particle arrival \citep{RichCane10}, confirming that the interplanetary medium is not disturbed.  Moreover, the proximity in longitude of the parent active region and the footpoint of the nominal Parker spiral is consistent with this result (Table~\ref{t-event}).

 \begin{figure}  
   \centering
   \includegraphics[width=0.5\textwidth, clip=]{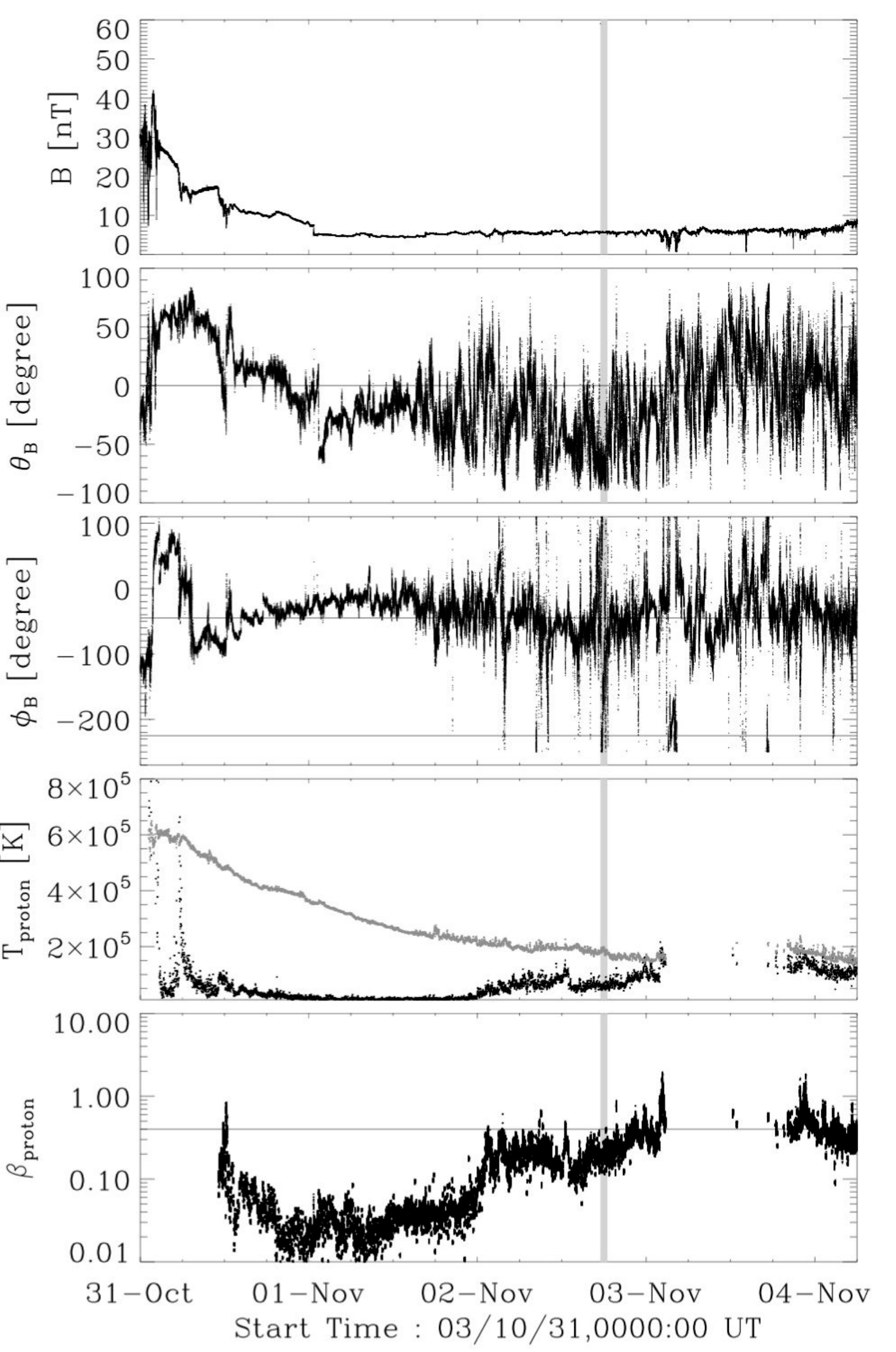}
   \caption{Interplanetary data around GLE~67: it is located in the back region of a magnetic cloud.  Same drawing convention as Figure~\ref{fig1}.}   
   \label{fig2}
 \end{figure}

\subsubsection{ICME, magnetic cloud or back region}
\label{s-imfmc}

For the five GLEs occurring on Jul. 14, 2000 (GLE~59), on Apr. 18, 2001 (GLE~61), on Oct. 28, 2003 (GLE~65), on Oct. 29, 2003 (GLE~66), and on Nov. 2, 2003 (GLE~67), we find that particles propagate in a transient magnetic structure, such as in the sheath of an ICME, in a magnetic cloud or in the back region of a magnetic cloud.

The GLE~on Nov. 2, 2003 is detected at Earth at 17:30~UT (Figure~\ref{fig2}). On that day, the magnetic field strength ($B \simeq 5~\rm{nT}$) and its longitude (fluctuating around $\phi_{B}\simeq -50~\degr$) are typical of a solar wind structured by the Parker spiral.  However, $\theta_{B}$ is not aligned with the ecliptic plan ($\simeq -50~\degr$) as one expects for a Parker spiral. Moreover, both $T_{\rm obs}< T_{\rm exp}$ and $\beta_p < 0.4$ are inconsistent with a quiet solar wind. Indeed, an ICME is detected, starting on Oct. 31, 2003 at 2:00~UT and ending on Nov. 2, 2003 at 00:00~UT. 

Within the ICME, from $\simeq$~8:00~UT on Oct. 31 to 1:00~UT on Nov. 1, both $\theta_{B}$ and $\phi_{B}$ have a relatively coherent global evolution. Moreover, $T_{\rm obs}< T_{\rm exp}/2$, implying that a magnetic cloud is present.  After 1:00~UT on Nov. 1,  $\theta_{B}$ and $\phi_{B}$ evolve progressively away from their values found at the rear of the magnetic cloud, and their fluctuation level increases. This behavior is characteristic of the back region of a magnetic cloud that displays intermediate properties between a magnetic cloud and the solar wind \citep{Dasso_al06, Dasso_al07}. 

According to \citet{RichCane10} the ICME ends when $\beta_p$ shows a sharp increase.
Nevertheless, even though $\beta_p$ increases, it remains less than $0.4$ and the ratio $T_{\rm obs}/T_{\rm exp}$ is still less than $0.5$ almost until the end of Nov. 2 (Figure~\ref{fig2}).  In addition, both $\theta_{B}$ and $\phi_{B}$ still display similar mean values as at the rear of the magnetic cloud.  Thus, the back region extends until the end of Nov. 2 (when $T_{\rm obs} \approx T_{\rm exp}$ and $\beta_{\rm p}\approx 0.4$).  We conclude that the energetic particles of GLE~67 propagate up to the Earth in the extended back region of the previous magnetic cloud.

Following a similar reasoning, we identified the magnetic structures of the IMF for the other GLEs (the analysis is reported in Section~\ref{A-MC} of the Appendix).  We conclude that relativistic particles related to the GLE~61 propagate in the sheath of an ICME (Section~\ref{A-GLE61}). The GLE~66 occurs inside a magnetic cloud \citep[previously studied by][]{Mandrini_al07}.  It is worth to notice here that this magnetic cloud is related to the CME ejected from the Sun on Oct. 28, 2003 at 11:30~UT during the solar eruption producing the GLE~65.  
Therefore, we conjecture that the relativistic particles produced during the flare/CME event on 29 Oct. 2003 have been injected in the footpoints of the CME launched on 28 Oct. 2003 (Section~\ref{A-GLE66}). For the GLE~59 and GLE~65, the temporal evolution of magnetic field and plasma parameters of the interplanetary medium displays intermediate properties between solar wind and magnetic cloud. A detailed analysis of all variables in a 4 days time interval around the GLE suggests that the particles related to GLE~59 and GLE~65 propagate in the back region (see Section~\ref{A-GLE59} and \ref{A-GLE65}).

The four GLEs (59, 61, 65, 66) are all poorly connected to Earth by a nominal Parker spiral (Table~\ref{t-event}). The solar eruptions associated to GLE~59, GLE~65 and GLE~66 are located near the central meridian, while the parent active region of GLE~61 is probably located behind the solar west limb. Thereby, the connection between the active region and Earth can not be ensured through the Parker spiral field line.  According to our analysis, we concluded that particles propagate in transient magnetic structures (ICME, magnetic cloud and back region). Since the coronal roots of such transient structures extend over a large longitudinal range, they can provide a magnetic path connecting the active region and the Earth. However, this does not imply that all GLEs associated with transient magnetic structures should be poorly connected. Indeed, the broad longitudinal extent of an ICME includes the well connected cases. An example is GLE~67 which is well connected by a nominal Parker spiral, while the high energy particles are traveling in a magnetic cloud back region.

 \begin{figure}  
   \centering
   \includegraphics[width=0.5\textwidth, clip=]{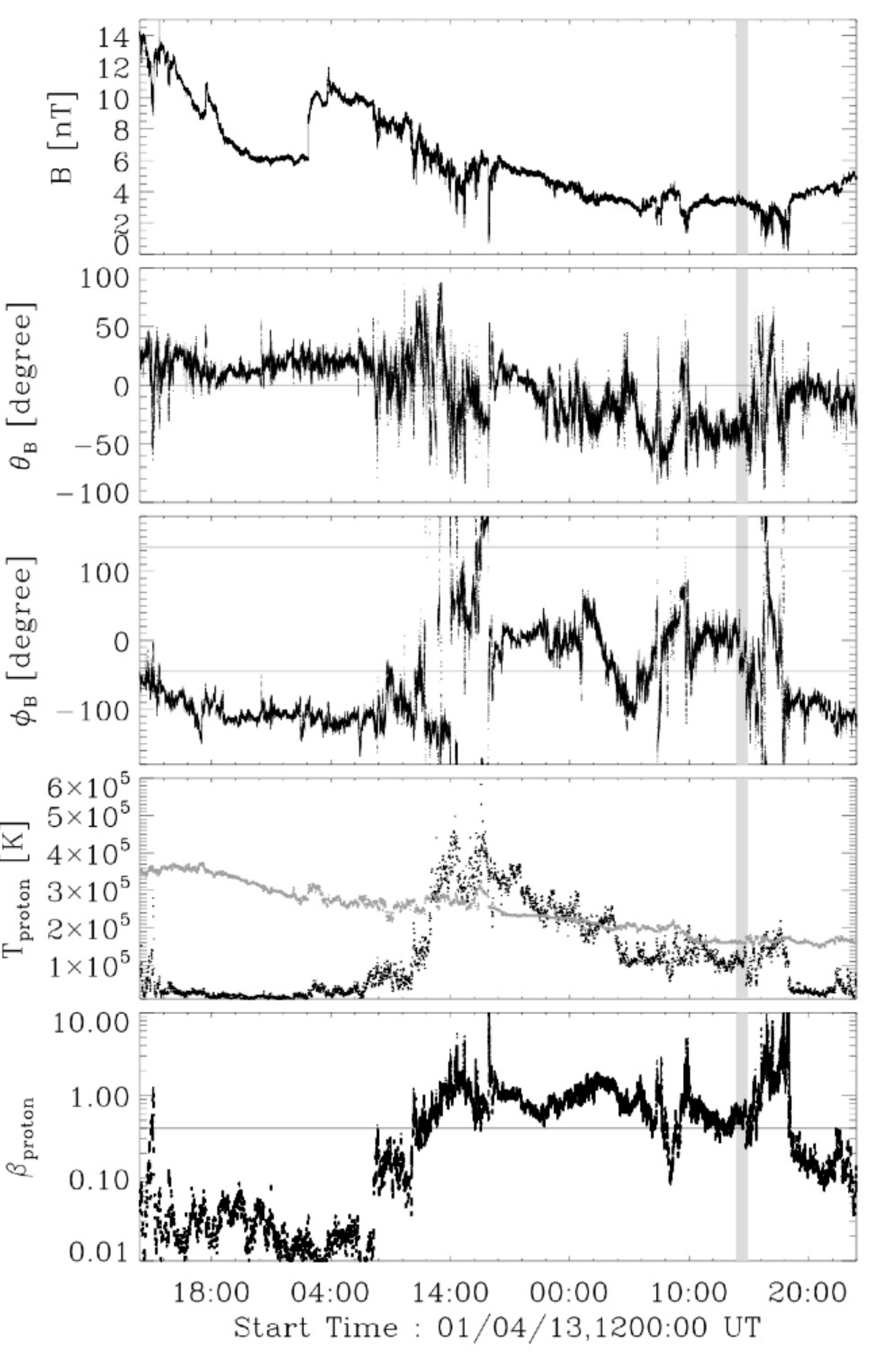}
   \caption{ Interplanetary data around GLE~60: example of a GLE~present in a SW disturbed by a previous ICME.  Same drawing convention as Figure~\ref{fig1}.}
   \label{fig3}
 \end{figure}

\subsubsection{Disturbed solar wind}
 \label{s-imfcompl}

The identification of the interplanetary magnetic structure of the GLE~60 and GLE~69 indicates that particles propagate in an interplanetary medium formed by a disturbed solar wind. 

We present in detail below the case of the GLE~60, occurring on  Apr. 15, 2001 (Figure~\ref{fig3}). The first relativistic particles impact the Earth at 13:55~UT. Between 10:00 and 16:00~UT, even though the magnitude of the magnetic field and the plasma parameters are close to typical solar wind values, the magnetic field orientation ($\theta_B \approx -45 \degr$ and $\phi_B \approx 0\degr $) does not correspond to the Parker spiral.  However, the  fluctuations of the magnetic field around fixed values of latitude and longitude are more characteristic of a magnetic structure associated with a quiet solar wind than with an ICME.  From the ICME list of \citet{RichCane10}, the last strong magnetic perturbation was an ICME, starting the Apr. 13 at 07:34~UT and ending on Apr. 14 at 12:00~UT, while the next ICME starts about 3 hours after the relativistic particles reached Earth.  Then, we suggest that particles propagate in a solar wind strongly disturbed by the preceding ICME. 

Similarly, the magnetic structure in which the relativistic particles of GLE~69 propagate is identified to be also a solar wind magnetic topology disturbed by a previous ICME (Section~\ref{A-GLE69}). 

Judging from their longitudes,  GLE~60 (W~85) and GLE~69 (W~55) are, respectively, poorly and well connected to the Earth by the Parker spiral (Table~\ref{t-event}). However, the apparently good connection  of the active region related to the GLE~69 does not imply that the magnetic field connecting the Sun to the Earth is a Parker spiral. The anisotropy of this GLE, with the strongest and fastest rise seen by neutron monitors in Antarctica, suggests a strong out-of-the ecliptic component of the magnetic field, which is actually observed  (Section~\ref{A-GLE69}).

\subsection{Summary}
 \label{s-imf-Summary}

Using ACE magnetic field and plasma parameter measurements, we identified the structure of the interplanetary magnetic field.  We show that, only in 3 events out of 10, relativistic particles possibly propagate in a solar wind structured by the Parker spiral, while in the 7 other events, relativistic particles propagate in a transient IMF (ICME, magnetic cloud, back region or disturbed solar wind). Among these 7 events, 5 have a solar source region significantly far (more than $30\degr$) from the theoretical  location of the footpoint of the Parker spiral at the solar source surface.
  
\section{The interplanetary length}
\label{s-vda}

 Previous studies of relativistic particles, measured at Earth by neutron monitors, assume that energetic particles travel along the Earth-connected Parker spiral field line. This assumption implies that particles that are not scattered travel roughly $1.1$-$1.3$ AU, depending on the solar wind speed.  In fact, from our above results, most GLEs do not propagate in quiescent solar wind. Thus, we expect that the interplanetary length travelled by energetic particles should differ from the one of the Parker spiral.

\subsection{Velocity dispersion analysis using the rising phase method (RPM)} 
\label{Principle-RPM}

A commonly used estimate of interplanetary travel paths of energetic particles is the velocity dispersion analysis (VDA) of the time when the first particles arrive at the detector. The method is based on the assumption that the first energetic particles are simultaneously released from a small acceleration site in the solar corona and propagate scatter-free in the interplanetary medium. Therefore, the less energetic particles arrive at the Earth later than the most energetic ones, implying a velocity-dependent time dispersion of the first particle detection. Plotting the arrival time of the first particles measured at Earth as a function of the inverse of particle speed ($1/v$), one predicts a linear relationship whose slope is a measure of the interplanetary distance travelled by the energetic particles, and whose intercept with the $1/v=0$ axis is the solar release time. 

\begin{figure*}
   \includegraphics[width=\textwidth,clip=]{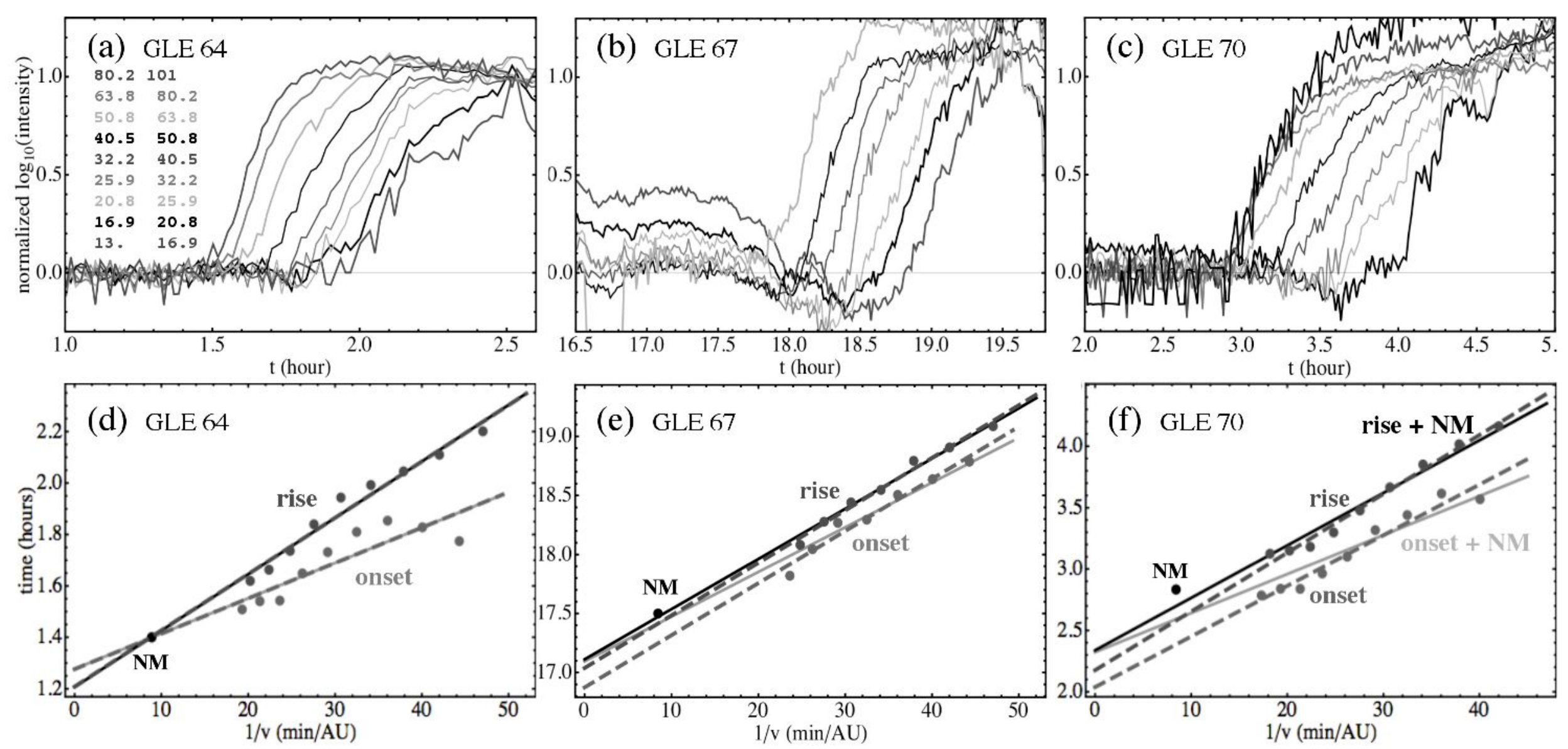}
\caption{Example of the RPM and VDA analysis. {\it top row}: normalized time profile  of the logarithm of the intensity during the rising phase without any time shift, for GLEs 64, 67 and 70, respectively on panels a, b and c (see Section~\ref{Principle-RPM}). The energy range (in MeV) of channels is added on the panel a. {\it Bottom row, panels d, e and f}: plots of rise and onset times in function of the inverse of the proton velocity. The dark grey (or blue) points correspond to the rise times determined by RPM apply to ERNE data (Section~\ref{Principle-RPM}), and the related linear fit is displayed by a dashed line (dark grey or blue). The linear fit related to the data set including the NM point (labeled and plotted in black or red), in addition to the previous ERNE points, is represented by a continuous line (black or red).  The medium grey (or green) points correspond to the onset times of ERNE data and the related linear fit is displayed with a dashed line and the same color.  The linear fit with the NM point added is shown with a continuous line and light grey (or pink). A color version is available in the electronic version. 
}
\label{fig-rpm}
\end{figure*}

The major practical problem of the method is the definition of arrival times of particles at the detector. The onset time is commonly defined as the instant when the intensity exceeds a given level above background. But this value is strongly affected by fluctuations in the individual energy channels, which may introduce a distortion and produce a large error both in the slope (the length) and the intercept (the release time) of the resulting plot. This is what we experienced when applying different such methods to the events under study. We therefore propose a new method, called the rising phase method (RPM).  The RPM is based on the same principle as the VDA, but it compares well-defined reference times during the rise phase of the time profiles at different energies, and thereby provides an estimate of the interplanetary length that is much less sensitive to background fluctuations than the classical VDA. Practically we proceed as follows:

 \begin{enumerate}
\item The intensity profile in each energy channel is divided by the background intensity averaged over one hour before any evident rise of the signal. Since the signal has a broad range of variation we take the logarithm of this ratio, making the background level zero. 
\item Depending on the energy spectrum of the event, the different energy channels reach different maximum values. In order to compensate this hardness effect, we normalize the above logarithm of fractional intensity by its maximum found just after the rise phase. This normalization is realized in a time interval typically between 10 and 20 min after the rise phase, with a time-shift function of the channel energy in order to compensate approximately the differential arrival time in the energy channels. Thus, the logarithm of this normalized intensity increases from 0 to 1 in the rising phase of the signal. Three examples are shown in the top row of Figure~\ref{fig-rpm}. 
\item The rising phase between the levels $0.2$ and $0.8$ is fitted by a straight line (assuming an exponential rise of intensity). The fit reduces the effects of the fluctuations superposed on the general increase of intensity, as seen in several curves in Figure~\ref{fig-rpm}, panels a \& c. We refer to the instant when the fitted straight line is at level 0.5 as the rise time at the corresponding energy. This corresponds to the maximum cross-correlation of the rising profiles.
\item Finally, we plot the rising parts of intensity profiles as a function of observing time minus the rise time, and control visually how well the profiles in the different energy ranges superpose. 
 \end{enumerate}

As the classical VDA, the RPM provides a set of times $t_{\rm rise}$ which depend on the energy of the channels, hence on the velocity of the detected energetic particles.  Then, the slope of a linear fit of  $t_{\rm rise}$ {\it versus} $1/v$ provides an estimation of the interplanetary length $D$ (bottom row of Figure~\ref{fig-rpm}). The estimation of $D$ depends only on the relative timing of the channels. 

The linear fit also provides an estimation of the solar release time, $t_{\rm SRT}$ of the particles (supposed to be independent of energy) as the intercept of the straight line with the $1/v=0$ axis. We have to keep  in mind that this is an upper limit, because it refers to a time during the rise of the intensity profile. We estimate the error on $D$ and $t_{\rm SRT}$ with the standard error of the fitted parameters.

We believe that the use of the rise times gives a more reliable determination of the travel path than using the onset times mainly for three reasons. As already stated, the method does not depend on the fluctuating background, since it uses the part of the signal well above it. Moreover, the fluctuations of the background and of the signal should have a minor influence since the method cumulates the information over a finite time interval (most of the rising part of the signal).  Finally, the rise time is computed in the steepest part of the rising phase of the time profile, where the timing is the most accurate. These three properties contrast with the onset time method which relies on a local measure above the fluctuating background where the signal just starts to increase. 

However, the time dispersion determined by the RPM may be affected by systematic biases, such as scattering during the propagation of particles from the Sun to Earth.  Since the mean free path increases with increasing particle energy \citep{Droge00}, the lower energy particles should arrive with an extra delay compared to the arrival time that one expects. This extra delay at low energy may lead to an increase of the length obtained by the RPM. The dependance of the RPM's results in respect to the particle scattering is hard to assess. However, we expect that one of the following three situations should apply: (1) scattering of the first arriving particles is negligible, and the slope of the onset time vs $1/v$ plot gives the geometrical path, {\it e.g.} GLE 63 \& GLE 64; (2) scattering dominates the propagation to the extent that the slope is mostly determined by the energy dependence of the scattering mean free path, {\it e.g.} GLE 70 ; (3) both the geometrical path and scattering shape the plot. Thereby, the RPM gives at least the apparent length travelled energetic particles, but this apparent length may be a mix of the geometrical length and the additional length created by the scattering.

\subsection{Application to the selected events}
\label{vda}

\subsubsection{Data set and event selection} 
\label{vda-Data}

We perform the velocity dispersion analysis, based on the RPM, using measurements of the High Energy Detector (HED) of the Energetic and Relativistic Nuclei Experiment (ERNE) instrument \citep{Torsti_al95} on board SoHO, located at the L1 Lagrangian point.  The energy range covered by ERNE/HED extends from $13~\rm{MeV}$ to $130~\rm{MeV}$.  Data are provided in 10 customized energy channels. The 5 highest energy channels, between $40-130~\rm{MeV}$, are usually distorted during GLEs (R.~ Valtonen, private communication): in order to privilege the detection of heavy ions, the onboard software raises the threshold for the detection of protons and He ions during large events. As a result, the proton intensities between $40-130~\rm{MeV}$ may display recurrent indentations. We correct this artifact in order to recover a regular time profile to which the RPM can be applied. Identifying the start and end times of the indentations by the sudden change of intensity, we apply a correcting factor to each interval by using the measured values obtained just before and after the jump. These corrections have a negligible effect on the fitted parameters. Still, they allow to use higher energy channels which are closer to the NM data. 

Since it is not clear {\it a priori} if the relativistic protons measured by the neutron monitors are just the high-energy extension of the SEP spectrum measured by ERNE or a distinct population, we perform separate velocity dispersion analyses with and without the data point of the GLE~as determined by \citet{Moraal_al11}. 
We use the mean energy of each ERNE channel to compute the velocity of energetic protons.
The RPM can not be applied to neutron monitor measurements, which respond to the integral proton spectrum above the local rigidity cutoff, as determined by the geomagnetic field configuration.

Only seven of the ten GLEs have suitable ERNE observations for the RPM:  there is no data for the GLE~59 and for the GLE~66, energetic particles of the SEP are smothered in a larger energetic particles flux probably accelerated at the bow shock of the magnetic cloud. Finally, we can neither use the ERNE data of the GLE~69 because the signal decreases and no bump has been detected in the time interval during which we expect the arrival of energetic particles related to the GLE.

\begin{table}
\caption[]{Path lengths and release times using RPM}
\label{t-length}
\centering
\begin{tabular}{  c  c  c  c c c c c}
\hline 
1                &  2  &   3        &     4     & 5  \\  
\# GLE& \multicolumn{2}{c}{Interplanetary length }&\multicolumn{2}{c}{Solar release time } \\
                     & $-$ NM, rise  & $+$ NM, rise  & $-$ NM, rise& $+$ NM, rise \\
\hline
60 &$ 1.50\pm 0.16$&$ 1.55\pm 0.10$& 13:45 $\pm 05$& 13:43$\pm 03$\\
61 &$ 1.61\pm 0.12$&$ 1.63\pm 0.09$& 02:23 $\pm 04$& 02:23$\pm 03$\\  
63 &$ 1.27\pm 0.06$&$ 1.39\pm 0.07$& 05:35 $\pm 02$& 05:31$\pm 02$\\
64 &$ 1.31\pm 0.09$&$ 1.32\pm 0.06$& 01:12 $\pm 03$& 01:12$\pm 02$\\
65 &$ 1.59\pm 0.16$&$ 1.89\pm 0.17$& 11:15 $\pm 05$& 11:05$\pm 05$\\
67 &$ 2.67\pm 0.15$&$ 2.55\pm 0.09$& 17:02 $\pm 05$& 17:07$\pm 03$\\
70 &$ 2.87\pm 0.13$&$ 2.56\pm 0.16$& 02:11 $\pm 04$& 02:20$\pm 05$\\
\hline
\end{tabular}
\begin{list}{}{}
\item[{\it Note :}]Effective interplanetary length (AU) travelled by  protons and the associated solar release time (UT). The error bars correspond to the standard deviation of the linear fit.
{\it Col.~1}: the label of the GLEs. 
{\it Col.~2}: length computed from the rise times and the mean energy of ERNE channels - without the NM's data. 
{\it Col.~3}: as Col.~2 but with the NM's point. 
{\it Col.~4}: the solar release time of energetic protons from the RPM applied to only to ERNE data (as Col.~2, see Figure~\ref{fig-rpm}). 
{\it Col.~5}: as Col.~4 but with the NM's point.
\end{list}
\end{table}

\begin{table}
\caption[]{Path lengths and release times using the classical VDA}
\label{t-lengthVDA}
\centering
\begin{tabular}{cccc}
\hline 
1   &      2        &         3     &      4      \\  

\# GLE&\multicolumn{3}{c}{Interplanetary length}   \\
	&$-$ NM, onset  & $+$ NM, onset & Reames       \\
	   \hline
60 &$ 0.93\pm 0.26 $&$ 0.85\pm 0.17 $&$ 1.59\pm 0.01$\\
61 &$ 1.60\pm 0.49 $&$ 1.34\pm 0.38 $&$ 1.80\pm 0.10$\\
63 &$ 0.78\pm 0.07 $&$ 0.76\pm 0.05 $&$ 1.64\pm 0.06$\\
64 &$ 0.83\pm 0.17 $&$ 0.82\pm 0.13 $&$ 2.16\pm 0.05$\\
65 &$ 3.12\pm 0.44 $&$ 2.41\pm 0.43 $&$ 1.38\pm 0.03$\\
67 &$ 2.65\pm 0.23 $&$ 2.27\pm 0.18 $&$ 2.01\pm 0.04$\\
70 &$ 2.47\pm 0.21 $&$ 1.91\pm 0.29 $&$ 2.81\pm 0.02$\\
    \hline
\end{tabular}
\begin{list}{}{}
\item[{\it Note : }] Effective interplanetary length (AU) travelled by  protons and the associated solar release time (UT). 
{\it Col.~1}: the label of the GLEs. 
{\it Col.~2}: length computed from the onset times and the maximal energy of ERNE channels - without the NM's point.  
{\it Col.~3}: as Col.~2 but with the NM's point. The error bars of Col.~2 and 3 correspond to the standard deviation of the linear fit, increased by the error on the determination of the onset time (due to background fluctuations).
{\it Col.~4}: length from \citet{Reames09}.
\end{list}
\end{table}

\subsubsection{Results}
 \label{vda-Results}

Figure~\ref{fig-rpm} displays the results of the RPM for three GLEs selected to illustrate the three kinds of interplanetary magnetic structures previously identified. Even though the intensity profiles display significant fluctuations (top panels of Figure~\ref{fig-rpm}), the RPM succeeds to derive the rise time of energetic particle fluxes in each energy channel.  When these rise times are plotted as a  function of $1/v$ (bottom panels of Figure~\ref{fig-rpm}), they display a linear relation. 
 
We report in Table~\ref{t-length} the interplanetary lengths resulting from our RPM with and without the NM data, respectively.  We found that for 5 out of 7 events, the difference between the two lengths computed with and without the neutron monitor data point is not larger than $\simeq 0.05~\rm{AU}$, so well inside the error bars. 

However, we emphasize that our RPM considers an instant near half maximum during the rise of the ERNE time profiles as the onset time, whereas the neutron monitor onset time is the instant when the first neutron monitor signal was detected. These times are not directly comparable. A number of the events of Table~\ref{t-length}, for which ERNE measured well-defined onset time profiles, are associated to weak GLEs, where the time when the signal starts to exceed background is probably a rough upper estimate of the actual start, and actually also designates a time during the rise of the GLE. It is likely that the systematic error of the neutron monitor onset time compensates to some extent the different timing definitions of the two data sets. Therefore we note that the timing is consistent, but we can not use this finding to draw firm physical conclusions on the simultaneity of proton acceleration from tens of MeV to GeV energies. 

\begin{figure}
   \centering
   \includegraphics[width=0.5\textwidth, clip=]{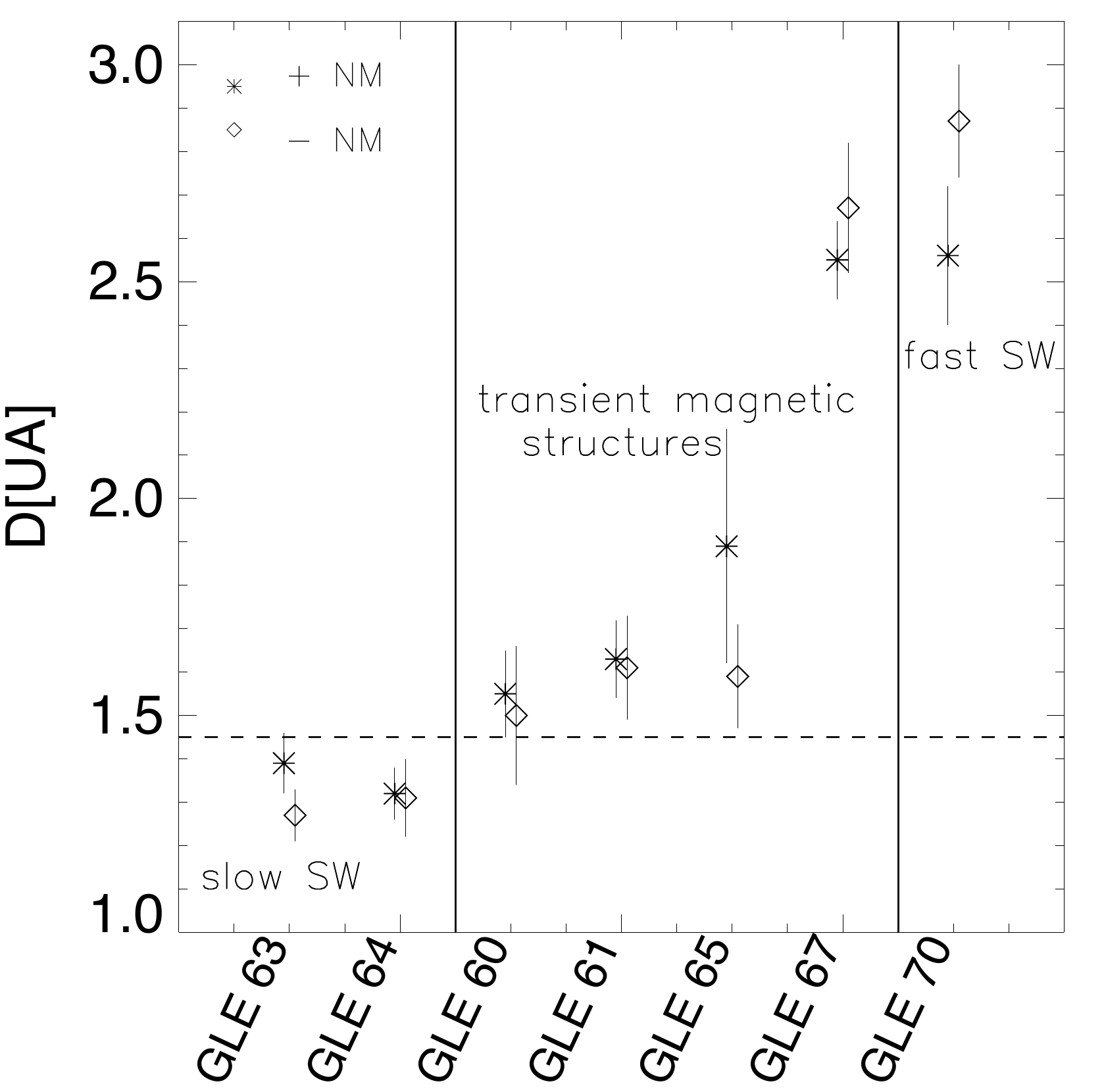}
\caption{Summary of the estimated length by the RPM, with (cross symbol) and without NM (diamond symbol), in relation with the magnetic topology.
}
\label{fig-correl-D}
\end{figure}

A discrepancy between the results with and without NM data is found in GLEs 65 and 70. They display a difference comparable to the error bar sizes (Table~\ref{t-length} and Figure~\ref{fig-correl-D}).
 For GLE~70 the results with and without NM has a length difference of $\simeq 0.31~\rm{AU}$. Irrespective of the exact numerical value, both analyses lead to an interplanetary path that is significantly longer than the Parker spiral for GLE~70. 

In the case of the GLE~65, we adapt the RPM. Indeed, the normalized logarithm of intensity has a temporal evolution significantly dependent of the energy channel and so has not a simple temporal shift as for other GLEs. We tested the usual normalization method, but it appears that the rising phases of the time profiles in different energy channels  are not parallel.  This difference in the time evolution of the channels indicates an evolution of the hardness of the spectrum during the rise phase. Indeed, one notices that a pre-increase appeared in the ERNE time profile in the high energy channels. According to \cite{Trottet_al08}, the acceleration of particles during this solar eruption displays a complex temporal structure, suggesting that several episodes of acceleration occur. Thereby, the pre-increase can result from a distinct acceleration episode than the one accelerating the particles of the main particle flux. 

In order to remove this effect for GLE~65, we use an earlier normalization, during the rising phase, at a time chosen such as to obtain rise time profiles in different energy channels that are parallel.  We also find that the rise times are more aligned with the neutron monitor data and that the lengths computed by the RPM with and without the NM point differ only by $\simeq 0.3~\rm{AU}$. 

\subsubsection{Comparison with results of classical VDA}
\label{ss-compare_vda}

We compare in this section the results of the RPM with the VDA method based on the estimation of the onset time, $t_{\rm onset}$, of the first particles in each channel. 
 Since the arrival time of the first protons is masked by the fluctuations of the background, we extrapolate the linear fit of the rise phase (see Section~\ref{Principle-RPM}) up to the mean level of the background. This defines the onset time, $t_{\rm onset}$.  The uncertainty on $t_{\rm onset}$ is partly due to the error on the linear fit of the rising phase (as for $t_{\rm rise}$), but it is also increased by the extrapolation toward the background level and by the background's fluctuations.  Considering the standard deviation of the background, this extra uncertainty on $t_{\rm onset}$ is typically of $3~\rm{min}$.  

Since the first particles are expected to have the highest energy, we compute $1/v$ from the maximum energy of each channel.  Using the maximum energy, $E_{\rm max}$, compared to the mean one, $E_{\rm mean}$ introduces a systematic difference in the estimated $D$ of a factor $1+\Delta E/(2E_{\rm mean}) \approx 1.06$ to the first order in $\Delta E/E_{\rm mean}$ (where $\Delta E=E_{\rm max}-E_{\rm min}$ and $E_{\rm mean}=(E_{\rm max}+E_{\rm min})/2$).

We find that the results with $t_{\rm onset}$ are typically much more fluctuating and incoherent with the neutron monitor data than the results with $t_{\rm rise}$ (Figure~\ref{fig-rpm}). Moreover, the length found with the onset times is unrealistically short (below 1~AU) for some GLEs (Table~\ref{t-lengthVDA}).

In Table~\ref{t-lengthVDA}, we also report the interplanetary length computed by \citet{Reames09} with a velocity dispersion analysis of the onset time of protons and heavy ions.  These results have some large differences with our VDA results (Table~\ref{t-lengthVDA}). However, Reames results are broadly consistent with our rise time results (Table~\ref{t-length}), except mainly for three events: GLEs~63, 64 and 67.  Even though the VDA has been applied to the heavy ion data (from WIND/EPACT/LEMT) in order to get a lower background level, the fluctuations of the signal before and during the rising phase, {\it see e.g.} Figure 3 in \citet{Reames09}, and the temporal resolution (5-10 min) introduce limits to the accuracy of the estimated lengths.

In summary, we conclude that RPM gives a more meaningful evaluation of the interplanetary length travelled by energetic particles than the VDA, whatever the data set is. This conclusion is further strengthened in the next sub-section.

\subsection{Comparison of the RPM results with the identification of the IMF structure}
\label{ss-lengthti}

\subsubsection{IMF structure and interplanetary path length}
\label{ss-lengthti-Relation}

From the comparison of the interplanetary path length inferred from the RPM (Table~\ref{t-length}) and the interplanetary structure identified in Table~\ref{t-event}, a clear relation can be established between the magnetic field along which energetic protons propagate and the distance they travelled. Figure~\ref{fig-correl-D} summarizes the interplanetary lengths computed without and with NMs  (Col.~2 and 3 in Table~\ref{t-length}) as a function of the IMF topology (Col.~8 in Table~\ref{t-event}). 

The protons in the rising phase of GLE~63 and GLE~64 that are associated to a Parker spiral magnetic structure, travelled approximately $1.3\pm 0.1$~\rm{AU}.  These interplanetary lengths are consistent with the Parker spiral magnetic topology that has a theoretical length of $\approx 1.15$~AU for a slow solar wind. These results differ from \citet{Reames09} who found a length of $\approx 1.6~\rm{AU}$ and $2.2$~\rm{AU} (Table~\ref{t-lengthVDA}) for GLE 63 and 64, respectively.  Within the error bars ($\pm 0.06$), these lengths are compatible with a Parker spiral only if an extra delay is present, so if a physical mechanism can provide it independently of the proton energy.

By contrast of GLE~63 and~64, the interplanetary lengths travelled by protons producing GLEs~60, 61, 65, and 67, are longer than $1.5~\rm{AU}$ (see Table~\ref{t-length}).  Such lengths are fully compatible with the magnetic topology of an IMF created by transient magnetic structures, such as an ICME, magnetic cloud, back region and even a disturbed solar wind.

The two lengths obtained with and without NMs, for GLE~65 (see Table~\ref{t-length}), are both consistent with the back region of a magnetic cloud (because a back region is formed by reconnection between the solar wind and the flux rope, see the end of Section~\ref{s-imf-Characteristic}). However, the longer length is supported by the results from \citet{Miroshnichenko_al05}. Studying this specific GLE, they showed that the delay of the proton arrival should be due to  the perturbation of the IMF and that protons should travel roughly $2~\rm{AU}$.  In addition, they estimated that this length is consistent with a simultaneous release of relativistic protons with relativistic neutrons, estimated at 10:56~UT from the NM data ($t_{SRT}=t_{NM}-D/v$).  From our RPM analysis, and for linear regression with and without NMs, the proton release occurs later (Table~\ref{t-length}).  However, the solar release time may have been overestimated because the pre-increase detected in the high energy channels of ERNE has been ignored in the determination of the time dispersion (Section~\ref{vda-Results}).

The last relativistic event of the solar cycle 23 (GLE~70) displays also a discrepancy between the magnetic structure and the interplanetary length.  Indeed, the length travelled by protons is $\approx 2.6$-$2.9$~AU,  while the IMF displays characteristics of the Parker spiral.  We note that this is the only GLE~of our sample that is detected while the Earth is in the fast solar wind. It is well known from {\it Ulysses} observations in the fast solar wind that SEP are more strongly scattered there than in the slow wind \citep{Snd-04}. Since the mean free path of the particles increases with increasing energy, particles of lower energy will be more strongly delayed than if they propagated scatter-free. This may contribute to an overestimation of the interplanetary path length. Moreover, the solar source of the fast solar wind is typically further  from active regions than the edge of the open flux region, so from the source region of a GLE.  Then, when energetic particles are detected in the fast solar wind they are expected to have been transported a long way from their acceleration site.  This extra delay could contribute to the long apparent length found for GLE~70. Such effect is expected to be smaller in ICME and slow solar wind cases as the guiding magnetic field is located closer to the GLE solar source.

Except for this complex GLE~70, the interplanetary length is consistent with the magnetic structure observed during the particle event.  We conclude that the propagation of energetic protons occurs in an IMF specific to each GLE, and frequently different from the Parker spiral often used. Moreover, this shows that the geometry of the magnetic structure affects the length travelled by particles from the Sun to the Earth. 

  \begin{figure}
   \centering
   \includegraphics[width=0.5\textwidth, clip=]{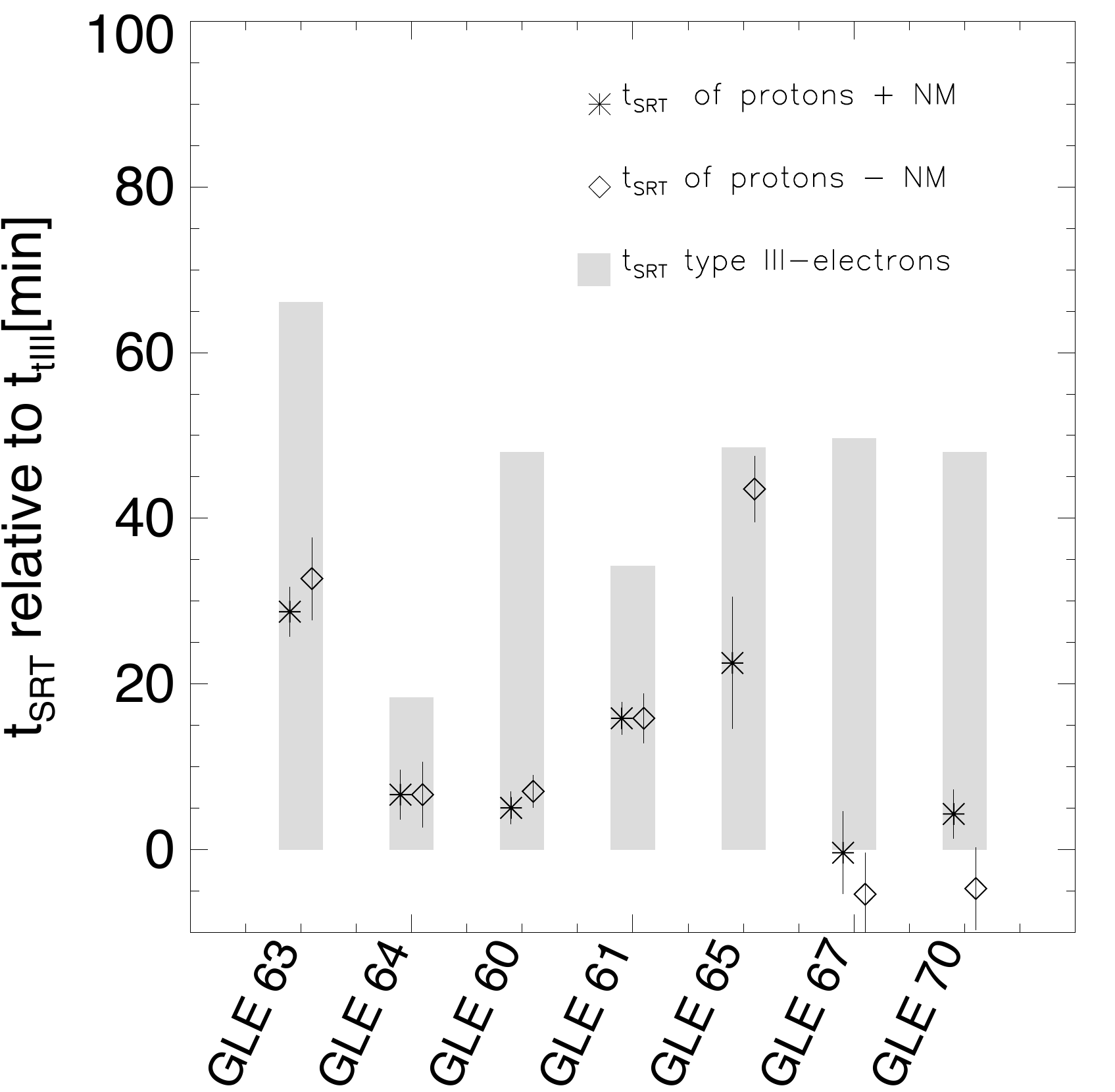}
  \caption{Comparison of the solar release time of relativistic protons obtained from the RPM with (cross symbol) and without NM (diamond symbol), with time interval of electrons injection as deduced from type III (grey shaded bars). The onset time of type III bursts is set as the origin of the vertical axis.
  }
 \label{fig-tSRT-TIII}
  \end{figure}

\subsubsection{When are protons injected into the IMF?}
\label{ss-lengthti-injection-time}

The propagation of non-thermal electrons in the interplanetary medium generates radio emission in the decametric wavelength range with a characteristic drift towards lower frequencies as the electrons propagate away from the Sun, towards lower ambient densities, along open magnetic field lines. These are type~III bursts. Large groups of type~III bursts accompany most major SEP events \citep{Cane_al02}.  At $14~\rm{MHz}$, it is generally assumed that the electromagnetic waves are emitted at $\sim 2-3~\rm{R_{\odot}}$ from the solar surface.  The timing of the type~III bursts at this frequency gives an indication of the time of electron injection into the high corona. Although the electrons are accelerated lower in the corona, the travel time from the acceleration site to the 14~MHz source is only a few tens of seconds, so this correction of time is negligible.

We computed the time interval of electron injection with the WIND/ WAVES \citet{Bougeret_al95} data at $14~\rm{MHz}$ and we compared them to the solar release time.  The Figure~\ref{fig-tSRT-TIII} synthesizes the temporal relation of the injection of the non-thermal electrons and the relativistic protons for the 7 studied events (Section~\ref{vda-Results}).

For 6 of the 7 events studied, the solar release time of energetic protons
is enclosed in the time interval of electron injection (Figure~\ref{fig-tSRT-TIII}).  For GLE~67 the solar release of energetic protons precedes that of electrons by $3-5~\rm{min}$. This is within the uncertainty of the method (see Table~\ref{t-length}). We therefore conclude that the first energetic protons during these events are released during the interval of the electron injection traced by the type~III bursts.

This timing consistency is an additional argument supporting our results on the magnetic topology of the interplanetary medium and its role for the propagation of energetic particles. Essentially, it gives a second proxy confirming that the interplanetary length has been well evaluated.

\section{Conclusion and Discussion} 
 \label{Conclusion}

We performed independent studies based on different methods, in order to define and constrain the characteristics of the interplanetary magnetic field during GLEs. The main results are summarized below.

First, in most of the events, relativistic protons propagate in transient magnetic structures and not along the Parker spiral field lines.

Second, the magnetic structure of the IMF is consistent with the location of the parent active region
of the GLE: when the {\it in situ} measurements show a Parker spiral, the parent active region is within $\pm30^\circ$ of the nominal footpoint, whereas it may be farther away when the IMF has a transient configuration.

Third, the effective interplanetary length computed through our velocity dispersion analysis (rising phase method, RPM) 
for energetic and relativistic protons is in agreement with the magnetic topology of the IMF deduced from {\it in situ} measurements. This length travelled is $\approx 1.3$~AU when the protons travel in the slow solar wind while it is in the range $1.5$-$2.6$~AU for transient magnetic structures. For the only GLE having protons traveling in the fast solar wind, the effective length is much longer than expected ($2.6$-$2.9$~AU rather than $1.1$~AU). This may indicate that an extra mechanism delayed the protons the more strongly, the lower their energy. Pitch angle scattering is a possible candidate, especially since it is expected to be stronger in the fast than in the slow solar wind.

Fourth, our results are consistent with the idea that the first protons producing SEPs and GLEs are accelerated/released simultaneously into the interplanetary medium, although they do not demonstrate this. The weakness of neutron monitor signatures in a number of events where the RPM could be applied to ERNE data precludes firm conclusions. The analysis also supports the idea that the first arriving energetic protons are not significantly scattered during their travel from the Sun to the Earth.

Fifth, according to the solar release time, energetic protons are injected during the injection of non-thermal electron beams producing type III bursts, supporting the above results on the impact of the interplanetary magnetic structure on the proton transport.

The common release of relativistic protons and type III-related electrons has already been suggested in a detailed study of GLE~69 \citep{Masson_al09a}. The event was not studied here, because the interplanetary magnetic field measurements were unreliable. But the result is fully consistent with the conclusions drawn from the seven GLEs of the present sample. Our observational study suggests that the common release of type~III bursts emitting electrons and the first escaping energetic and relativistic protons is a general property of GLEs. 

In addition, energetic particles related to active region poorly connected to the Earth through the Parker spiral are may  connected through a transient magnetic structure. This result implies to reconsider the driven-shock acceleration being invoked to explain the detection of particles at Earth in poorly connected SEP events. 

 Finally, energetic protons propagating in a transient magnetic structure travel a longer distance than they should do in the Parker spiral case. This delay, due to the geometrical length of the interplanetary magnetic field, contributes to understand the quasi-systematic delay measured between the GLE~and the first radiative signature of accelerated particles at the Sun \citep{Cliver_al82}. Moreover, this longer length travelled by relativistic protons modifies the relation between the coronal radiative signatures and the detection at Earth of these protons. As an example, \citet{Li_al09} assumed that relativistic protons of GLE~70 travelled $1.1$~AU between the acceleration site and the Earth, leading them to conclude that particles have been flare-accelerated. However, our study clearly demonstrates that for GLE~70, one can not consider such a short path length of the interplanetary magnetic field. 

\begin{acknowledgements}
{\bf The authors gratefully acknowledge R. Wimmer-Schweingr\"uber and A. Balogh  for helpful discussions on the contents of this paper.}
This research has made use of NASA's Space Physics Data Facility (SPDF).
We thank the ACE instrument teams and their respective principal investigators, namely N.F. Ness (ACE/MAG) and D.J. McComas (ACE/SWEPAM).
The authors acknowledge financial support from ECOS-Sud
through their cooperative science program (N$^o$ A08U01).
This work was partially supported by the Argentinean grants:
UBACyT 20020090100264 (UBA), PIP 11220090100825/10 (CONICET), and PICT-2007-00856 (ANPCyT).
S.D. acknowledges support from the Abdus Salam International Centre for Theoretical Physics (ICTP),
as provided in the frame of his regular associateship.
S.D. is member of the Carrera del Investigador Cien\-t\'\i fi\-co, CONICET.
The work of S.M. is funded by a fellowship of D\'el\'egation G\'en\'erale 
pour l'Armement (DGA). 
{\bf S. M. thank the NASA Postdoctoral Program at the Goddard Space Flight Center,  administered by Oak Ridge Associated Universities through a contract with NASA for financial support.}
Financial support by the European Commission through the FP6 SOLAIRE Network (MTRN-CT-2006-035484) is gratefully acknowledged.
 We thank E. Valtonen for his very helpful assistance with the ERNE data.

\end{acknowledgements}


\begin{appendix} 

\section{Interplanetary magnetic structures for other GLEs} 
  \label{Appendix}

 \begin{figure}  
    \centering
   \includegraphics[width=0.5\textwidth, clip=]{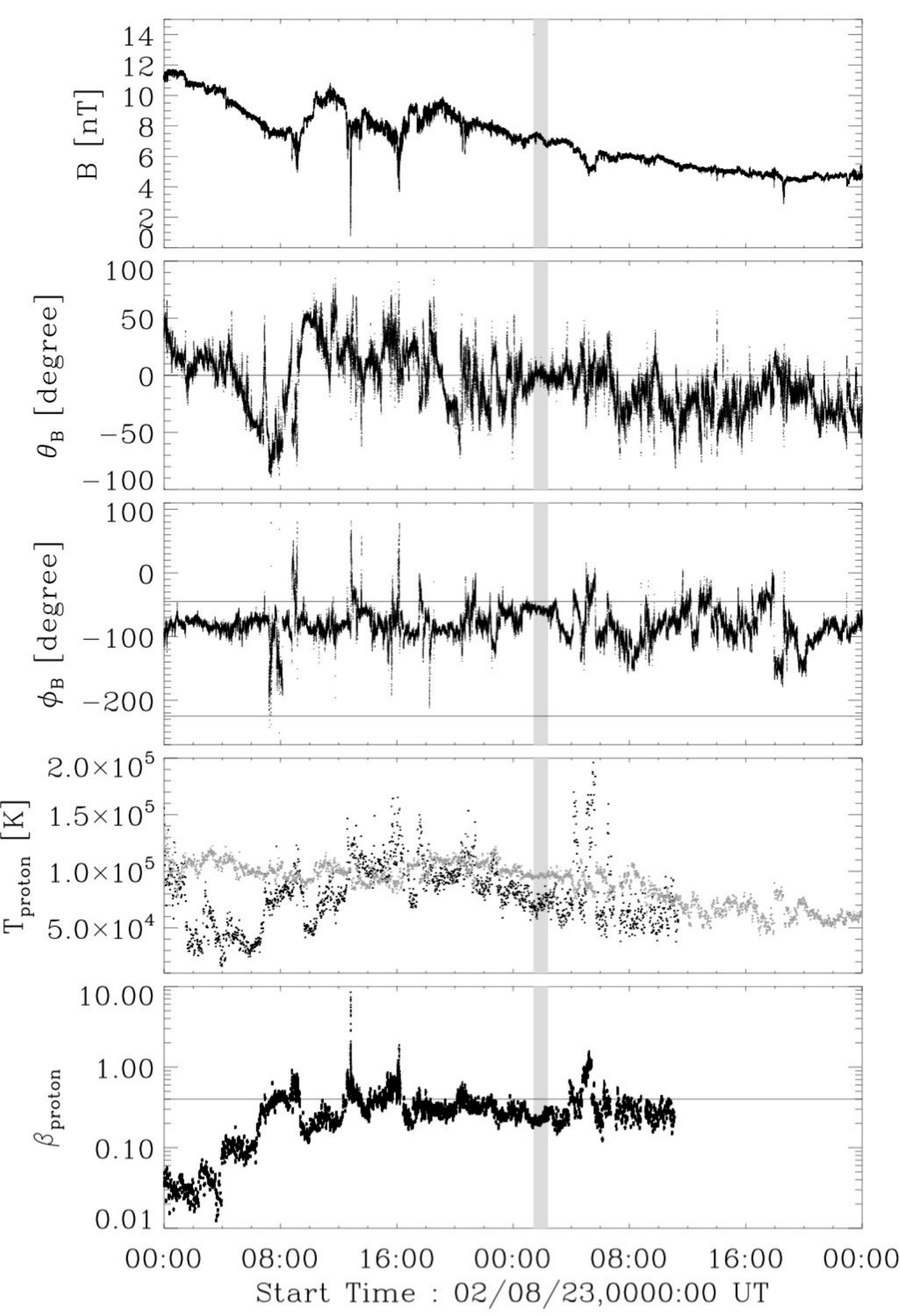}
   \caption{Interplanetary data around 
   GLE~64: it is located in a Parker like solar wind.  Same drawing convention as Figure~\ref{fig1}.}
   \label{figA1-1}
 \end{figure}

This appendix provides the analysis of the interplanetary structures associated with those GLEs which have not been
detailed in the main text.

\subsection{Solar Wind} \label{A-Solar-Wind}

\subsubsection{GLE~64 on Aug. 24, 2002} \label{A-GLE64}
 Relativistic protons of the GLE~64 arrive at Earth at 01:24~UT on Aug. 24, 2002, during the decrease of $B$ from $12$~nT to $5$~nT in two days (Figure~\ref{figA1-1}).
This decrease could suggest that protons travelled in the wake of the previous ICME ending on Aug. 21, at 14:00~UT \citep{RichCane10}. However, the orientation of the magnetic field $\theta_B \sim 0\degr$, $\phi_B =-45\degr$, the criteria on the temperature, $T_{\rm exp}\simeq T_{\rm obs}$, and $\beta_p \simeq 0.4$, are consistent with the Parker spiral magnetic field.

 \begin{figure}  
   \centering
   \includegraphics[width=0.5\textwidth, clip=]{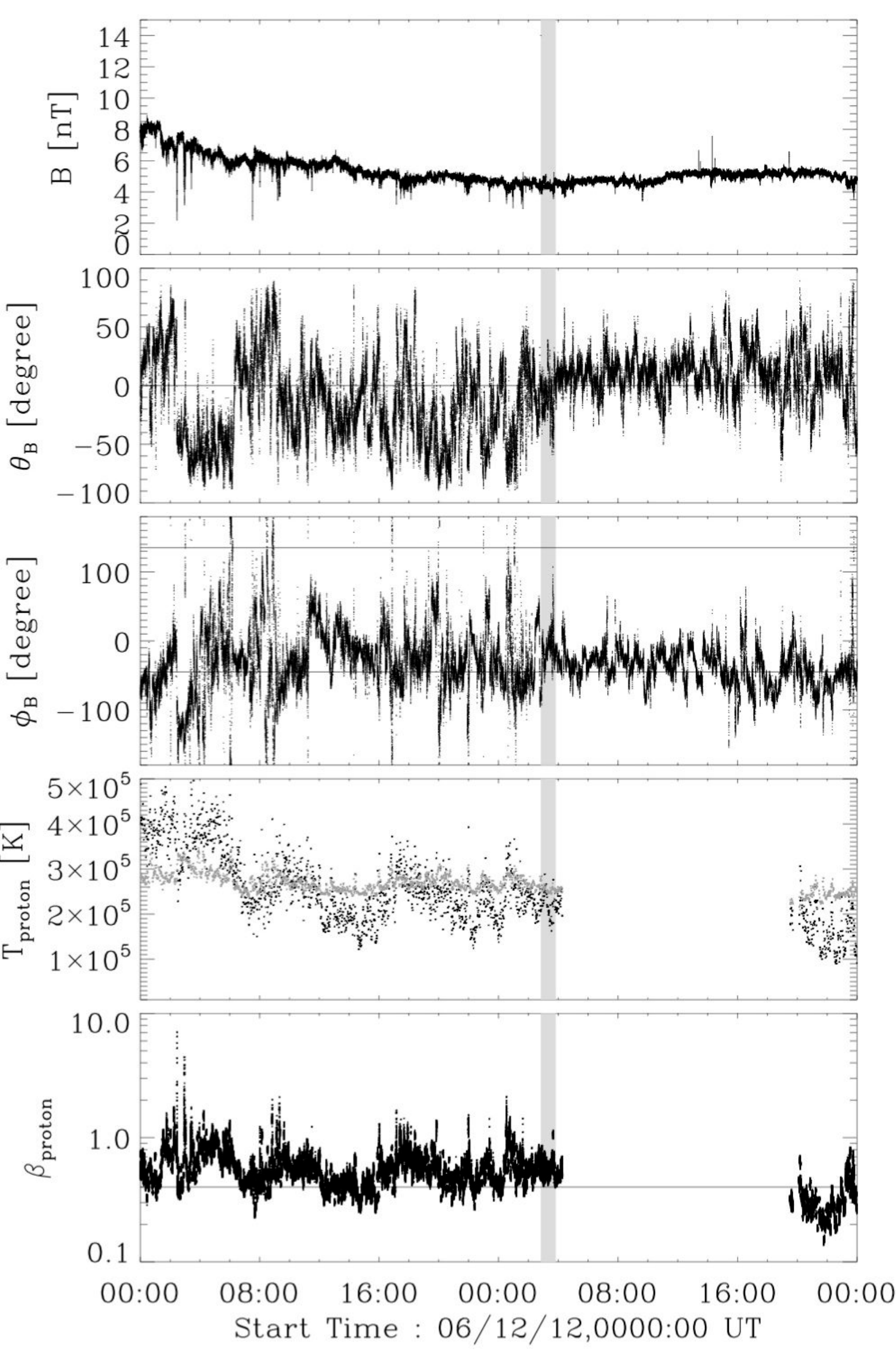}
   \caption{Interplanetary data around GLE~70: it is located in a Parker like solar wind.  Same drawing convention as Figure~\ref{fig1}.}
   \label{figA1-2}
 \end{figure}

\subsubsection{GLE~70 on Dec. 13, 2006}  \label{A-GLE70}

The magnetic field does not show any increase of magnitude or magnetic rotation, between 00:00~UT on Dec. 12 and 24:00~UT on Dec. 13 (Figure~\ref{figA1-2}).  Indeed, the magnetic field is non-coherent and fluctuates around $\theta_{B}=0~\degr$ and $\phi_{B}=-45~\degr$, with stronger fluctuations before the GLE~and weaker after that. 
Moreover, the expected temperature is roughly equal to the observed temperature, and $0.4 \la \beta_p \la 1$. The magnetic and plasma properties strongly suggest that the Parker spiral is the magnetic structure of the IMF, guiding the relativistic protons of the GLE~70, detected at the Earth at 02:50~UT on Dec. 13, 2006.

\subsection{ICME, magnetic cloud, or back region}  \label{A-MC}

 \begin{figure}  
    \centering
   \includegraphics[width=0.5\textwidth, clip=]{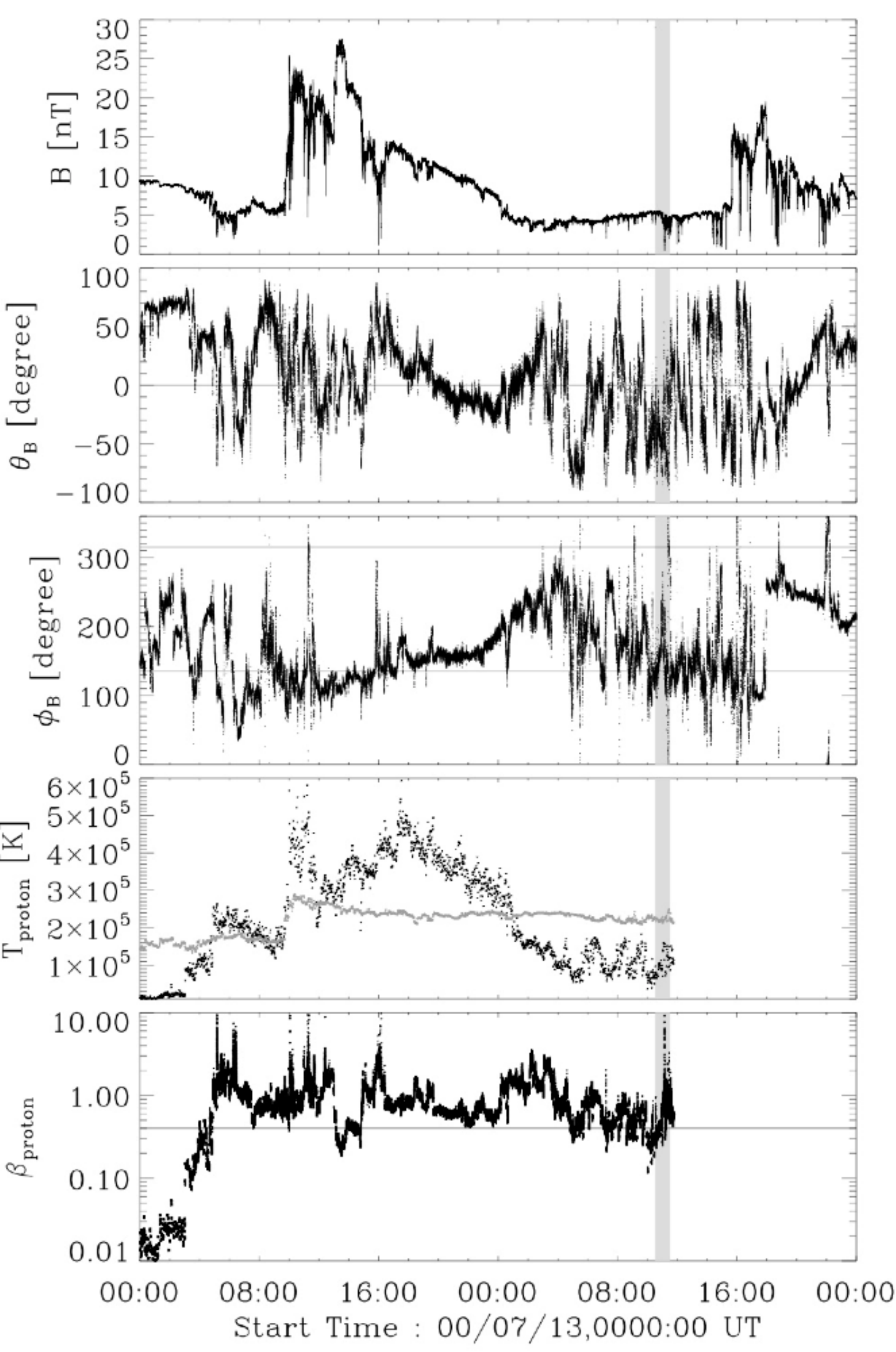}
   \caption{Interplanetary data around GLE~59: it is located in the back region of a hot flux rope.  Same drawing convention as Figure~\ref{fig1}.}
   \label{figA2-1}
 \end{figure}

\subsubsection{GLE~59 on Jul. 14, 2000}   \label{A-GLE59}

The magnetic field on Jul. 14, 2000 between 02:00~UT and 15:00~UT, enclosing the arrival time of the GLE~59 detected at 10:31~UT, does not display any common features of the well defined interplanetary magnetic field (Parker spiral, ICME, or magnetic cloud, Figure~\ref{figA2-1}). The magnetic field is weak ($B \simeq 3-5~\rm{nT}$), while the magnetic vector orientation is highly fluctuating and so does not have the typical $\theta_B$ and $\phi_B$ values of the Parker spiral. The plasma parameter evolution before the GLE also displays inconsistency: $T_{\rm obs} \simeq 0.5 T_{\rm exp}$, like in an ICME, whereas  $0.4<\beta_p<1 $, more typical of the solar wind. 

Before this time period, the magnetic field magnitude increases and its magnetic vector rotates coherently from Jul. 13, 2000 at 13:00~UT up to Jul. 14 at 02:00~UT, which is typical of a magnetic cloud with its sheath in front. Meanwhile, $T_{\rm obs}>T_{\rm exp}$ indicates that this is a hot flux rope. Even being hot, it was previously classified as a magnetic cloud \citep{Huttunen_al05,RichCane10}. These previous studies set the ICME end on Jul. 14, 2000 at 15:00~UT. The coherent rotation of the field, so the flux rope, ends at $\approx$02:00~UT.  Later on the magnetic field orientation has large fluctuations. Still, globally $\phi_{B}$ progressively evolves from its value at the rear of the flux rope ($\phi_B \approx 250-280\degr$) to an outward sector value. Thus, we propose that the region between 02:00~UT and 15:00~UT is the magnetic cloud back region ($\phi_B \approx 135\degr$). The magnetic field and plasma properties are intermediate between the solar wind and magnetic cloud.
Therefore, our analysis suggests that relativistic protons of GLE~59 arrived at Earth in the back region of a hot flux rope (magnetic cloud-like).

 \begin{figure}  
    \centering
   \includegraphics[width=0.5\textwidth, clip=]{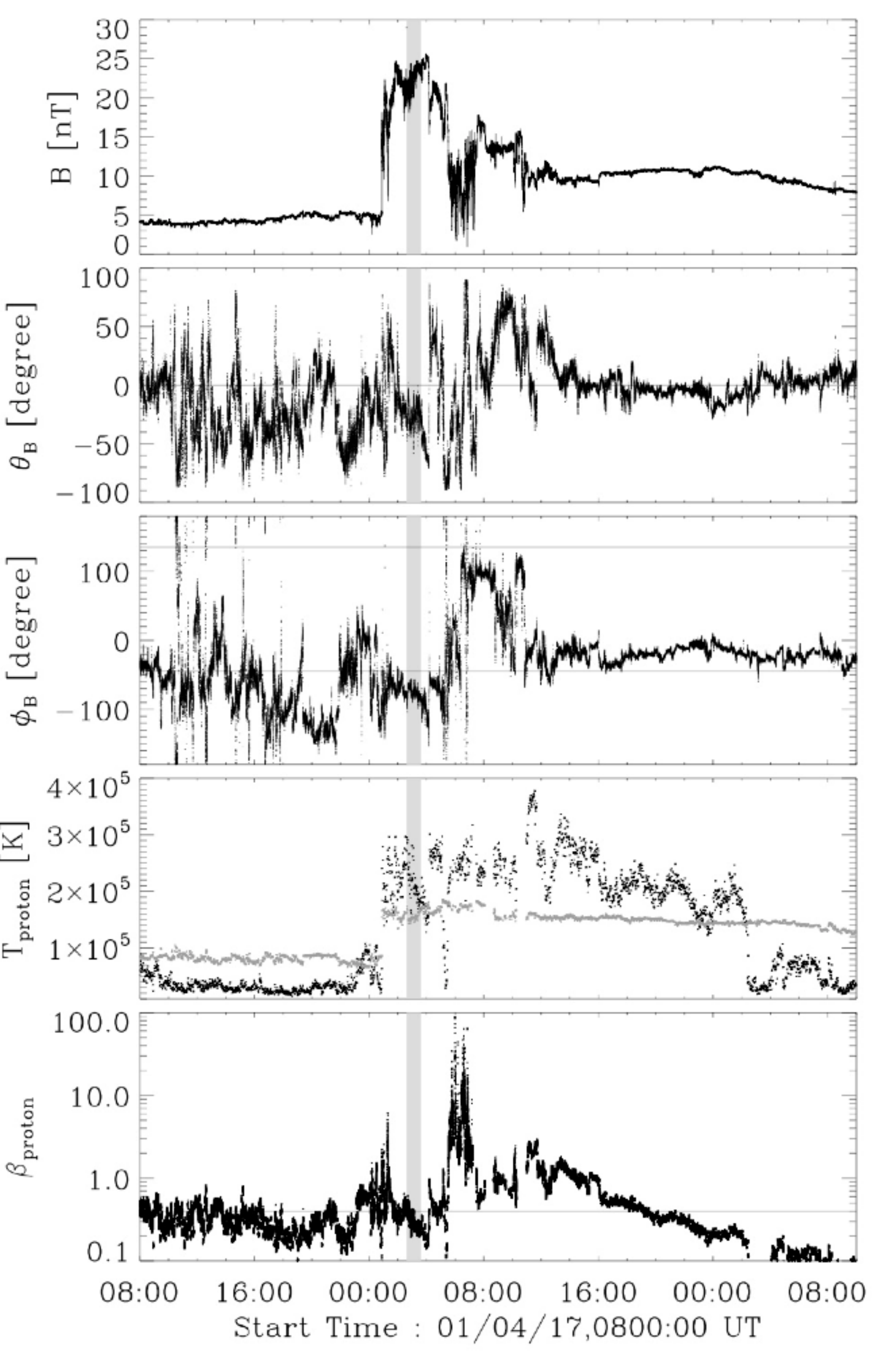}
   \caption{Interplanetary data around GLE~61: it is located in the sheath of an ICME.  Same drawing convention as Figure~\ref{fig1}.}
   \label{figA2-2}
 \end{figure}

\subsubsection{GLE~61 on Apr. 18, 2001}  \label{A-GLE61}

Relativistic protons on Apr. 18, 2001 (GLE~61), reach Earth at 02:36~UT, during a strong increase of the magnetic field strength ($B \simeq 20-25~\rm{nT}$, Figure~\ref{figA2-2}). After this period,  we identify an ICME ($B\sim10~\rm{nT}$, decreasing progressively to $\sim 5~\rm{nT}$) from the Apr. 18 at 12:00~UT to the Apr. 20 at 11:00~UT. This temporal evolution of {\bf B} is the signature of an ICME, preceded by its sheath. Our identification is in agreement with the boundaries determined by composition analysis \citep{RichCane10}.  However, this ICME is a particular case where the proton temperature is higher than the expected one and therefore $\beta_p\simeq 1$ in the front part of the ICME (Figure~\ref{figA2-2}).  We conclude that the relativistic protons of GLE~61 traveled in the sheath of this ICME.

 \begin{figure}  
   \centering
   \includegraphics[width=0.5\textwidth, clip=]{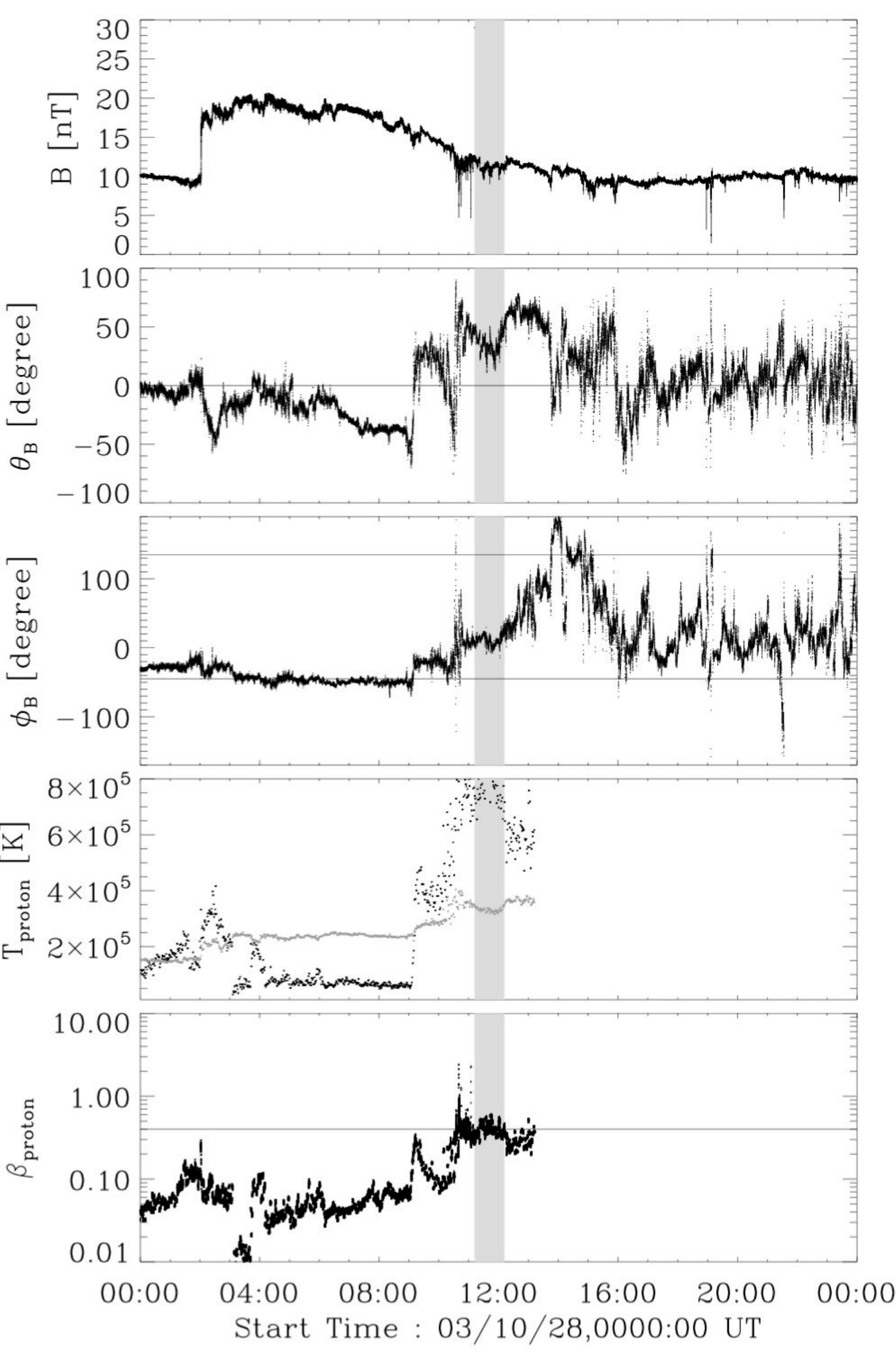}
   \caption{Interplanetary data around GLE~65: example of a GLE~present in the back of a magnetic cloud.  Same drawing convention as Figure~\ref{fig1}.}
   \label{figA2-3}
 \end{figure}

\subsubsection{GLE~65 on Oct. 28, 2003}  \label{A-GLE65}

 Figure~\ref{figA2-3} displays the evolution of the interplanetary magnetic field and plasma parameters 
for the GLE~65. The increase and the temporal evolution of the magnitude of the magnetic field starting at 02:00~UT on Oct. 28, 2003, the high coherence of the magnetic field vector, the ratio of the expected and the observed temperatures approaching 2 and the low value of $\beta_p < 0.4$ clearly indicate that ACE crosses an ICME. \citet{RichCane10} determined its end at 09:00~UT on Oct. 28.  They do not classify it as a magnetic cloud probably because the magnetic field rotation is only $\approx 60 \degr $.  This limited rotation indicates that the flux rope was crossed by the spacecraft with a large impact parameter (i.e. far from its axis).  Then, we conclude that the in-situ data are compatible with a magnetic cloud.

Around 09:00~UT the magnetic vector has a strong discontinuity, followed by many others until 16:00~UT on Oct. 28, as well as a progressive evolution of $\phi_B$. During this time interval, the observed temperature becomes about twice higher than the expected temperature.  The evolution of the magnetic field and of plasma parameters after 09:00~UT on Oct. 28 are compatible with the mixed properties of a hot back region in the wake of the magnetic cloud \citep{Dasso_al06, Dasso_al07}.  However, because the impact parameter of the magnetic cloud is large, the back region is also not observed in favorable conditions for a clear identification.  Finally, relativistic protons are detected at Earth at 11:12~UT, and we thereby conclude that  they propagate in this hot back region.

 \begin{figure}  
   \centering
   \includegraphics[width=0.5\textwidth, clip=]{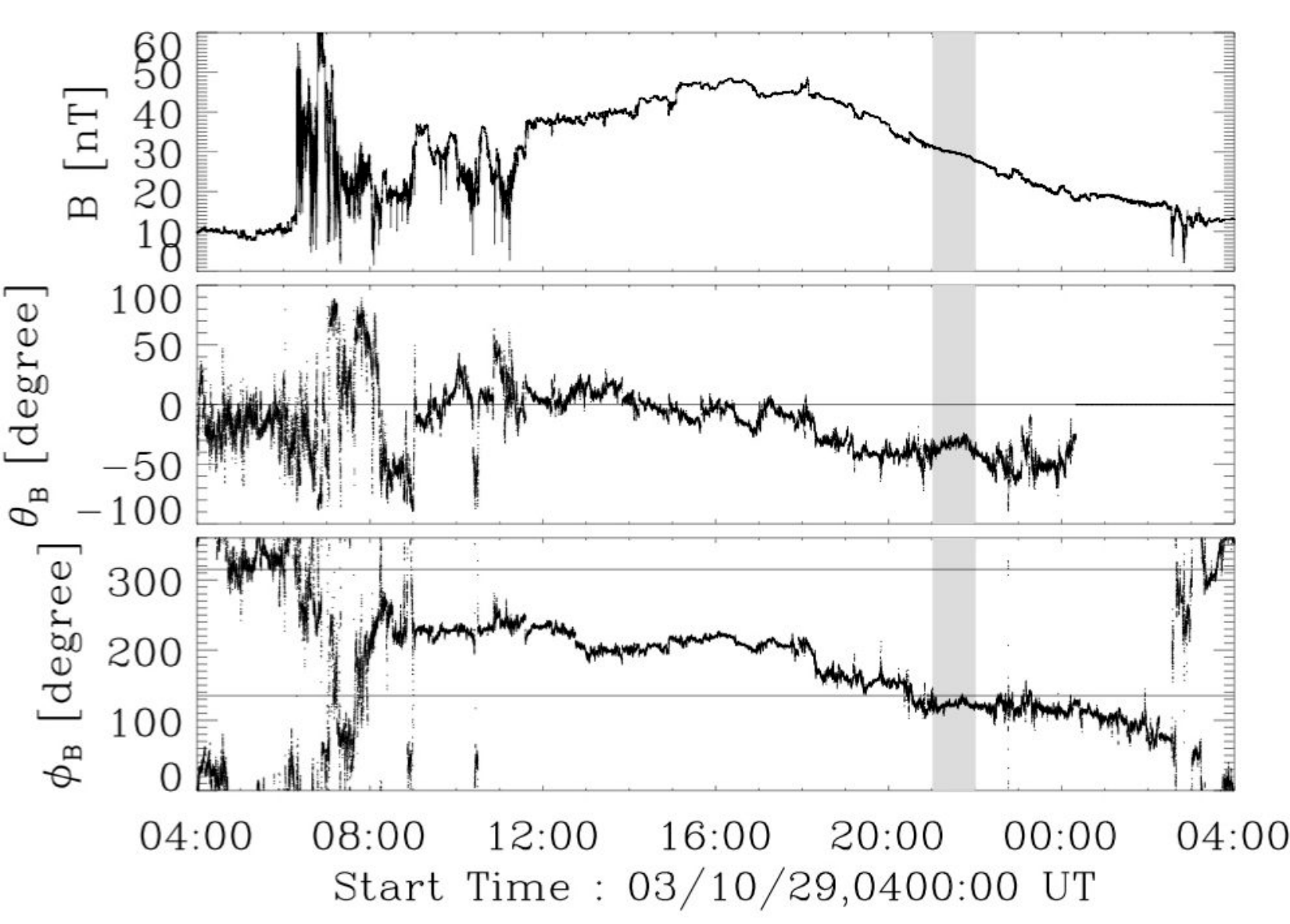}
   \caption{Interplanetary data around GLE~66: it is located inside a magnetic cloud. 
   Same drawing convention as Figure~\ref{fig1}. Plasma data are not available for this event. }
   \label{figA2-4}
 \end{figure}

\subsubsection{GLE~66 on Oct. 29, 2003}  \label{A-GLE66}

The interplanetary magnetic structure on Oct. 29, 2003 has already been identified as a magnetic cloud \citep{Mandrini_al07}. The magnetic cloud starts at 11:25~UT on 29 Oct. and ends at 02:00~UT on 30 Oct. The relativistic particles of the GLE~ reached Earth at 21:01~UT on Oct. 29, then they propagate in this well-identified magnetic cloud (Figure~\ref{figA2-4}). In addition, this magnetic cloud is related to the CME ejected from the Sun the 28 Oct. 2003 at 11:30~UT during the solar eruption producing the GLE~65.  Therefore, the relativistic protons produced during the flare/CME event the 29 Oct. 2003 have been injected in the foot-points of the CME launched the 28 Oct. 2003.

\subsection{Disturbed solar wind}  \label{A-Disturbed-SW}

 \begin{figure}  
   \centering
   \includegraphics[width=0.5\textwidth, clip=]{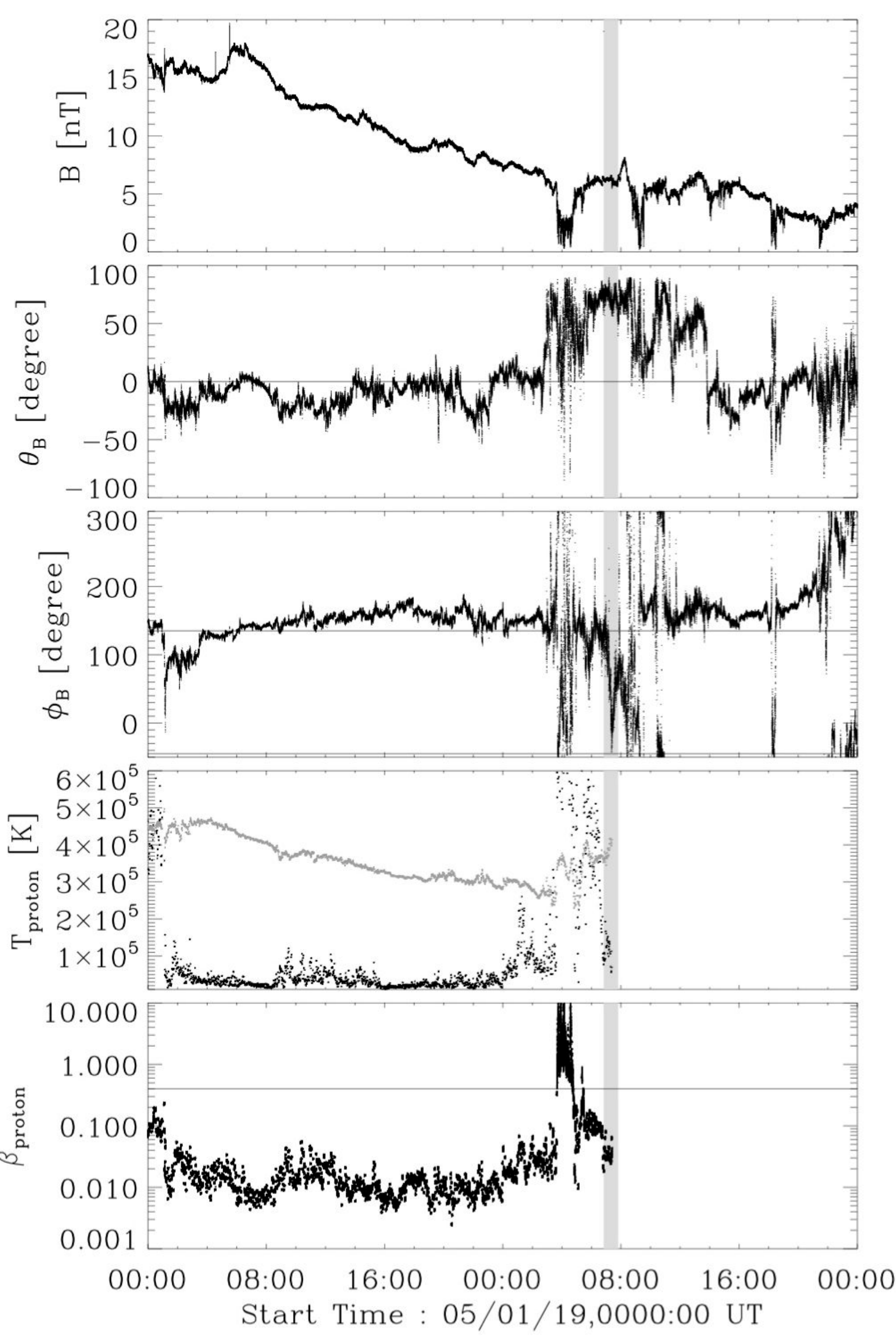}
   \caption{Interplanetary data around GLE~69: it is located in a disturbed solar wind.  Same drawing convention as Figure~\ref{fig1}.}
   \label{figA3-1}
 \end{figure}

\subsubsection{GLE~69 on Jan. 20, 2005}   \label{A-GLE69}

Before the GLE~69 on Jan. 20, 2005 at 6:49~UT, 
the interplanetary magnetic structure is identified as an ICME, starting on Jan. 18, at 23:00~UT and ending on Jan. 20, at 03:00~UT \citep{RichCane10}. During this time interval, the magnetic field has a significant strength (6-18 nT) with small fluctuations, and with a weak coherent rotation. Moreover, $T_{\rm obs}< 0.5 T_{\rm exp}$ and $\beta_p \simeq 0.02$.  So all the characteristics defining a magnetic cloud, except a large rotation of the field, are present.  This is a magnetic cloud-like case (Figure~\ref{figA3-1}). 

After 03:00~UT, the magnetic field and plasma parameters behavior do not correspond to the Parker spiral case as the magnetic field is nearly orthogonal to the ecliptic plane and $\phi_B$ is variable (so not characteristic of a solar wind sector).  The large rotation of the magnetic field around 03:00~UT would be an extreme case for the start of a back region, difficult to explain by the reconnection of magnetic cloud with a solar wind magnetic field, then we cannot identify it to a back region.  Still, the field nearly orthogonal to the ecliptic plane, the increase of the temperature and the $\beta_p$ are consistent with a disturbed solar wind located in the wake of the preceding ICME. Even though any clear magnetic structure can be identify, one can at least certainly conclude that relativistic protons related to the GLE~69 propagate in a disturbed solar wind.\\

\end{appendix}


{\large \bf Next page: Color version of one figure (for the electronic version). }
\newpage

\setcounter{figure}{3}
\begin{figure*}
   \includegraphics[width=\textwidth,clip=]{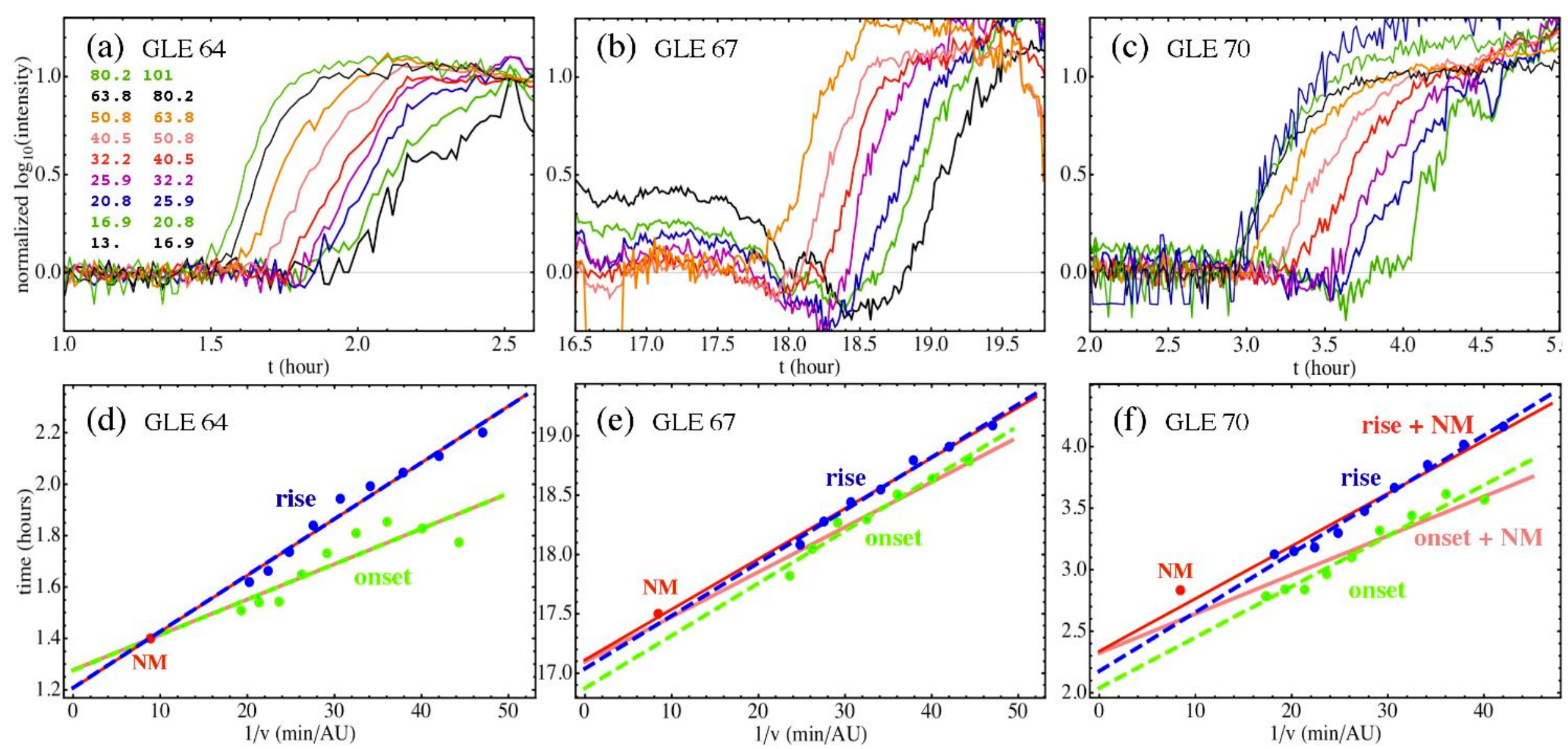}
\caption{Example of the RPM and VDA analysis. {\it top row}: normalized time profile  of the logarithm of the intensity during the rising phase without any time shift, for GLEs 64, 67 and 70, respectively on panels a, b and c (see Section~\ref{Principle-RPM}). The energy range (in MeV) of channels is added on the panel a. {\it Bottom row, panels d, e and f}: plots of rise and onset times in function of the inverse of the proton velocity. The dark grey (or blue) points correspond to the rise times determined by RPM apply to ERNE data (Section~\ref{Principle-RPM}), and the related linear fit is displayed by a dashed line (dark grey or blue). The linear fit related to the data set including the NM point (labeled and plotted in black or red), in addition to the previous ERNE points, is represented by a continuous line (black or red).  The medium grey (or green) points correspond to the onset times of ERNE data and the related linear fit is displayed with a dashed line and the same color.  The linear fit with the NM point added is shown with a continuous line and light grey (or pink). A color version is available in the electronic version. 
}
\end{figure*}

\end{document}